\documentclass[reprint,superscriptaddress,amssymb,amsmath,aps,showpacs,10pt,floatfix,prb,longbibliography]{revtex4-2}
\def\GCG{GdCu$_2$Ge$_2$}
\def\GCA{GdCuAl$_3$}
\def\cm{cm$^{-1}$}

\usepackage{graphicx}%
\usepackage{color}
\usepackage{epstopdf}
\usepackage{nicefrac}
\usepackage{lmodern}
\usepackage{amssymb}
\usepackage[T1]{fontenc}
\usepackage{booktabs}
\usepackage{amsmath}
\usepackage{amsfonts}
\usepackage[note-name=, use-sort-key = false]{notes2bib}

\usepackage{xcolor}

\usepackage{color}
\usepackage[colorlinks,bookmarks=false,citecolor=darkblue,linkcolor=red,urlcolor=blue]{hyperref} 
\definecolor{darkred}{rgb}{0.7,0.0,0.0}

\definecolor{darkblue}{rgb}{0,0.02,0.45}

\definecolor{darkgreen}{rgb}{0.02,0.45,0.0}

\definecolor{violet}{rgb}{0.8,0.2,0.6}

\begin{document}
\title{Electronic structure of the Gd-based intermetallics \GCG\ and \GCA }

\author{M. Pinteri\'{c}}
\affiliation{1. Physikalisches Institut, Universit{\"a}t Stuttgart, 70569 
Stuttgart, Germany}
\affiliation{Faculty of Civil Engineering, Transportation Engineering and Architecture, University of Maribor, 2000 Maribor, Slovenia}

\author{M. Dressel}
\affiliation{1. Physikalisches Institut, Universit{\"a}t Stuttgart, 70569 
Stuttgart, Germany}

\author{P. Puphal}
\email{p.puphal@fkf.mpg.de}
\affiliation{Max-Planck-Institute for Solid State Research, Heisenbergstraße 1, 70569 Stuttgart, Germany}

\author{M. Wenzel}
\email{maxim.wenzel@pi1.physik.uni-stuttgart.de}
\affiliation{1. Physikalisches Institut, Universit{\"a}t Stuttgart, 70569 
Stuttgart, Germany}

\date{\today}

\begin{abstract}
We present a temperature-dependent reflectivity study of single crystals of the ternary intermetallic compounds \GCG\ and \GCA\ over a broad spectral range (100--18000~\cm, equivalent to 12~meV--2.23~eV) down to 13~K. Below 2000~\cm, the optical spectra are dominated by the response of itinerant charge carriers exhibiting two distinct scattering rates. While the response of the highly-damped charge carriers shows negligible temperature dependence, the weakly-damped carriers closely follow the dc resistivity and are significantly suppressed in \GCA, consistent with its higher resistivity. Supported by density-functional-theory calculations, we further show that elemental substitution from \GCG\ to \GCA\ tunes the X-site $p$-orbital character (X = Ge, Al) of the low-energy electronic structure, while preserving key features such as linearly dispersing bands and saddle points despite the accompanying structural modifications. The reduced experimental plasma frequency of both compounds signals enhanced electronic correlations, which induce moderate band-energy renormalization.
\end{abstract}

\maketitle

\section{Introduction}
$f$-electron-based intermetallic compounds form a remarkably rich platform in condensed matter physics, where subtle changes in chemical composition can significantly alter both the magnetic and electronic properties of the system~\cite{Lai2022, Shatruk2019, Johrendt1997, Szytula1989, Tan2018, Petit2015, Pfleiderer2009, Dzero2010, Puphal2020, Haussermann2002}. Within this broad material landscape, ternary compounds derived from the layered tetragonal BaAl$_4$ structure provide a vital playground for studying the interplay between localized 4$f$ magnetic moments and itinerant $s$-, $p$-, and $d$-electrons. Based on the stacking sequence, these materials crystallize in three different space groups including the centrosymmetric $I4/mmm$ group (ThCr$_2$Si$_2$ structure), and $P4/nmm$ (CaBe$_2$Ge$_2$ structure) groups, and the non-centrosymmetric $I4mm$ space group (BaNiSn$_3$ structure), with broken inversion symmetry~\cite{Shatruk2019, Parthe1983}. These intermetallic ternary compounds display a wide variety of intriguing phenomena, such as unconventional superconductivity~\cite{Khim2021, Nogaki2021, Yuan2003, Schuberth2016, Steglich2023, White2015}, heavy fermion physics~\cite{Kimura2012, Anand2017, Lefevre2022, Gupta1983, Knebel1999}, non-Fermi liquid behavior~\cite{Mihalik2009, Gegenwart1998, Lai2022}, density-waves~\cite{Lee2020, Mydosh2020}, as well as topologically non-trivial electronic states~\cite{Ivanov2021, Kundu2022, Khim2021}.

Due to their half-filled 4$f$ shell, Gd-based ternary intermetallics feature strongly localized 4$f$ electrons and thus provide ideal systems to study these physical phenomena in the absence of additional complications such as Kondo effects, Jahn-Teller distortion, or crystal-electric-field effects \cite{Mallik1998, Guttler2016, Garcia2025, Kimura2021}. Dominated by RKKY interactions, these compounds typically exhibit well-defined magnetic ordering temperatures, displaying a large variety of magnetic ground states ranging from conventional ferromagnetism to helical and modulated antiferromagnetic orders~\cite{Mallik1998, Pottgen2010, Kliemt2017, Barandiaran1988, Szytula1989, Kumar2007}. Recently, magnetic-field-induced emergent skyrmion lattices have been reported in the non-centrosymmetric EuNiGe$_3$~\cite{Singh2023, Matsumura2024} (triangular skyrmion lattice), as well as in centrosymmetric GdRu$_2$Si$_2$~\cite{Yasui2020, Wood2023} and Gd$_2$PdSi$_3$~\cite{Bouaziz2022, Kurumaji2019} (square skyrmion lattice), both in the absence of geometrical frustration, and with negligible or even vanishing Dzyaloshinskii-Moriya interaction in the latter cases.

\begin{figure}
	\centering
	\includegraphics[width=1\columnwidth]{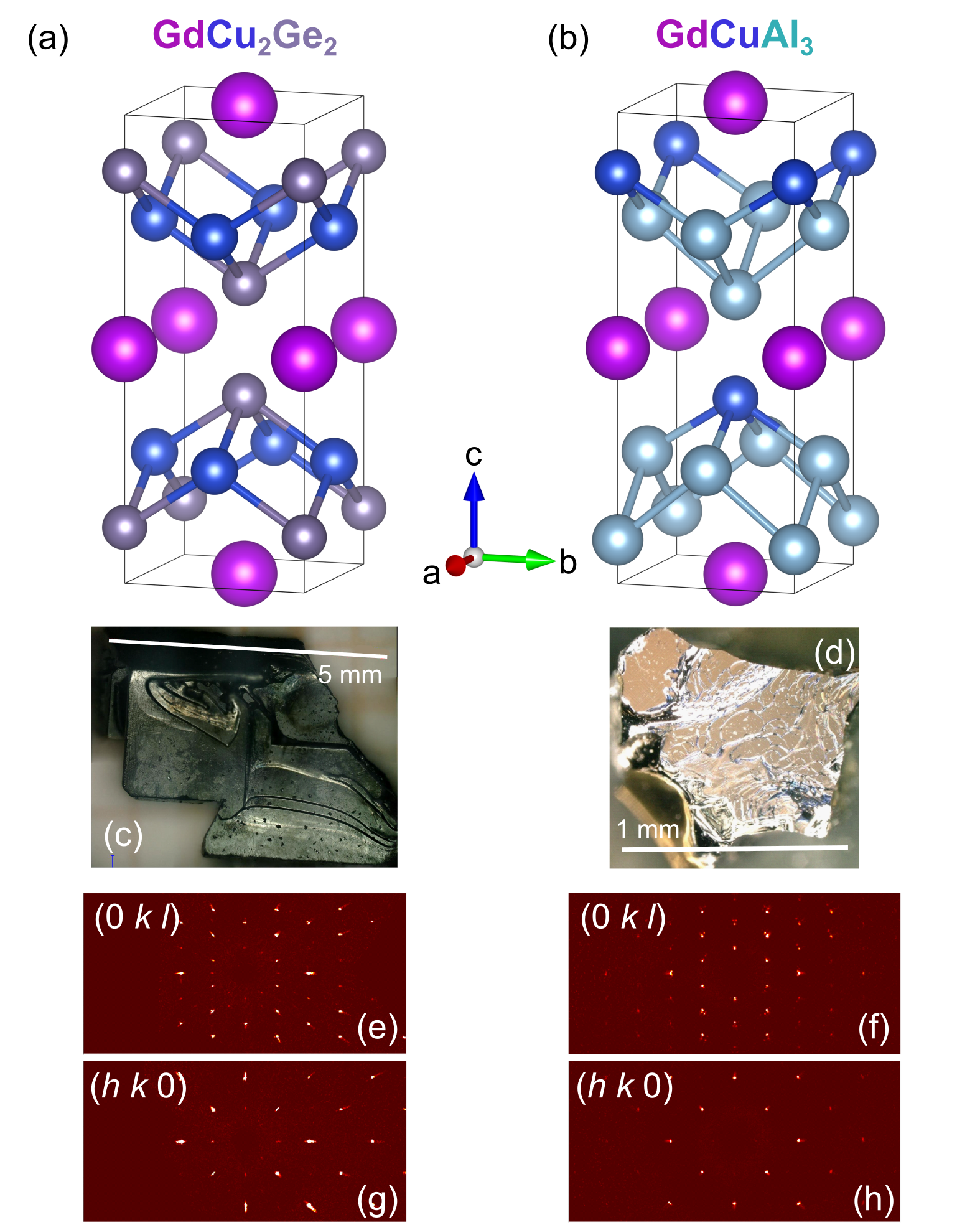}
	\caption{(a, b) Crystal structures of \GCG\ and \GCA, respectively, visualized with \texttt{VESTA}~\cite{Momma2008}. (c, d) Photographs of the single crystals used in the optical experiments prior to polishing. Backscattered x-ray Laue images of the same crystals are given in panels (e-h), confirming the tetragonal symmetry. }		
	\label{structure}
\end{figure}

Here, we address the modifications in the electronic structure of the ternary intermetallic compounds \GCG\ and \GCA\ arising from elemental substitution and, in the latter case, the loss of centrosymmetry [see crystal structures in Fig.~\ref{structure}(a) and (b)]. Using Fourier-transform infrared spectroscopy as a bulk-sensitive probe together with density-functional-theory (DFT) calculations, we unveil an intraband response with two distinct scattering rates and demonstrate robust band topology with tunable orbital character in ternary compounds derived from the layered tetragonal BaAl$_4$ structure.

\section{Methods}
\subsection{Synthesis and characterization}
\begin{table*}
\centering
\caption{Structural parameters extracted from single crystal 
refinement. GdCuAl$_3$ crystallizes in \textit{I4mm} with lattice 
constants of $a=b=4.1412(3)$~\AA, $c=10.6232(16)$~\AA, with a goodness 
of $R_{\mathrm{all}}=0.0490$, $wR_{\mathrm{gt}}=0.1208$. 
GdCu$_2$Ge$_2$ on the other hand crystallizes in \textit{I4/mmm} with 
lattice constants $a=b=4.0591(11)$~\AA, $c=10.233(3)$~\AA, and a 
goodness of fit of $R_{\mathrm{all}}=0.0487$, 
$wR_{\mathrm{gt}}=0.0941$. Note, that $U_{23}=U_{13}=U_{12}=0$.}
\begin{tabular}{lccccccc}
\hline
Atom & $x$ & $y$ & $z$ & $U_{\mathrm{eq}}$ & $U_{11}$ & $U_{22}$ & $U_{33}$ \\
\hline
Gd1 & 1 & 1 & 0.57795(5) & 0.0125(6) & 0.0103(6) & 0.0103(6) & 0.0169(9)  \\
Cu2 & \nicefrac{1}{2} & \nicefrac{1}{2} & 0.4503(3) & 0.0138(13) & 
0.0126(17) & 0.0126(17) & 0.016(3)  \\
Al3 & \nicefrac{1}{2} & 0 & 0.3281(8) & 0.0131(11) & 0.015(7) & 
0.012(7) & 0.013(2) \\
Al4 & \nicefrac{1}{2} & \nicefrac{1}{2} & 0.6704(11) & 0.015(3) & 
0.015(4) & 0.015(4) & 0.016(6)  \\
\hline\hline
Gd1 & 1 & 1 & \nicefrac{1}{2} & 0.0090(4) & 0.0112(5) & 0.0112(5) & 
0.0046(7)  \\
Ge2 & \nicefrac{1}{2} & \nicefrac{1}{2} & 0.61955(16) & 0.0101(4) & 
0.0138(5) & 0.0138(5) & 0.0027(8)  \\
Cu3 & 0 & \nicefrac{1}{2} & \nicefrac{3}{4} & 0.0120(5) & 0.0155(6) & 
0.0155(6) & 0.0050(10)  \\
\hline
\label{sc}
\end{tabular}
\end{table*}
High purity ingots of Gd, Cu and Al were cut and weighed according to a 1:1:3 stoichiometry in a glove box. The elements were then transferred into a fused silica ampule, which was sealed under vacuum. The ampule was then placed in a vertical tube 
furnace and heated from room temperature with 100~°C/h to 1190~°C, where the temperature was held for 2~h. After this, a slow cooling with 5~°C/h to 900~°C was started. The elongated boule was cracked and shiny crystals of typically 1~mm were extracted in air.

For \GCG, an attempt of GdCuGe$_3$ Sn flux was started by mixing 
0.22361~g Gd, 0.09036~g Cu, 0.30988~g Ge, and 3.37614~g Sn, in a glove box and 
loading the mixture into a Canfield crucible set of two crucibles separated by a 
sieve. These were sealed in a fused silica in vacuum. Afterwards the 
ampule was heated to 1000~°C with 100~°C/h and the temperature was 
held for 24~h. The furnace was slowly cooled at 1~°C/h to 550~°C, at 
which point the Sn flux was centrifuged through sieves. The ampules were 
opened in air and crystals of Gd$_2$CuGe$_6$ and GdCu$_2$Ge$_2$ with sizes of several millimeters were obtained.

The single crystals were characterized by single crystal XRD. We found 
the systems Gd-Cu-Ge to crystallize in the famous ThCr$_2$Si$_2$-type structure (space group \textit{I4/mmm}) with a corresponding stoichiometry, while GdCuAl$_3$ crystallizes in the famous GdNiSn$_3$-type structure (space group \textit{I4mm}). The corresponding Fourier maps are shown in Fig.~\ref{structure}(e-h) and the single crystal refinement results are summarized in Tab.~\ref{sc} (see Supplemental Material at \cite{SM} for CIF files). The crystals were oriented during this process and polished for optical characterization.

Four-point resistivity measurements were performed on the same crystals used in the optical experiments [see Fig.~\ref{structure}(c) and (d)]. The low dc resistivity values shown in Fig.~\ref{ROC}(a) and (b) indicate the highly metallic nature of these samples. The kinks observed around $T_{\mathrm{N}} = 10$~K and 12~K in the resistivity curves of \GCG\ and \GCA, respectively, correspond to the antiferromagnetic transitions, consistent with previous Mössbauer spectroscopy studies~\cite{Mulder1993, Mulder1994}.

\subsection{Optical measurements}
Temperature-dependent reflectivity measurements were performed on optically polished samples in the isotropic $ab$-plane down to 13~K ($T > T_{\mathrm{N}}$), covering a broad frequency range from 100 to 18000 \cm\ (12~meV -- 2.23~eV). For the high-energy range ($\omega\,>\,600\,\mathrm{cm^{-1}}$) a Bruker Vertex 80v spectrometer coupled with a Hyperion IR microscope was used, while the low-energy range ($\omega\,<\,600\,\mathrm{cm^{-1}}$) was measured with a Bruker IFS113v spectrometer and a custom-built cryostat. Freshly evaporated gold mirrors served as reference in these measurements. The absolute value of the reflectivity was obtained by an in-situ gold-overcoating technique in the far-infrared range, as described in Ref.~\cite{Homes1993}.

Below 100 \cm, we use standard Hagen-Rubens extrapolations, considering the metallic nature of our samples, while for the high-energy range we utilize x-ray scattering functions to extrapolate the data~\cite{Tanner2015}. The optical conductivity is then calculated from the measured reflectivity via Kramers-Kronig analysis~\cite{Dressel2002}.

\subsection{Computational}
Density-functional-theory (DFT) calculations of the band structure and optical conductivity were performed in the \texttt{Wien2K}~\cite{wien2k,Blaha2020} code with the Perdew-Burke-Ernzerhof flavor of the exchange-correlation potential~\cite{pbe96}. Experimentally determined structural parameters given in Tab.~\ref{sc} were used in the calculations. Fully relativistic self-consistent calculations were converged on the $12\times 12\times 12$ $k$-mesh. For the non-magnetic calculations presented in the main text, the open-core approximation was employed, in which the Gd 4$f$ electrons were treated as core states. Alternatively, a Hubbard $U = 10$~eV was added to the 4$f$ shell of Gd within the DFT+$U$ framework, using the FLL (fully localized limit) double-counting correction to push the minority 4$f$ states well above the Fermi level. In this approach, an in-plane antiferromagnetic order of the Gd atoms was imposed, necessary to stabilize the Gd$^{3+}$ ground state (see Supplemental Material~\cite{SM}). The optical conductivity was calculated using the \texttt{OPTIC} module~\cite{Draxl2006} on a denser $k$-mesh with up to $100\times 100\times 100$ points.

\section{Results and Discussion}
\begin{figure}
	\centering
	\includegraphics[width=1\columnwidth]{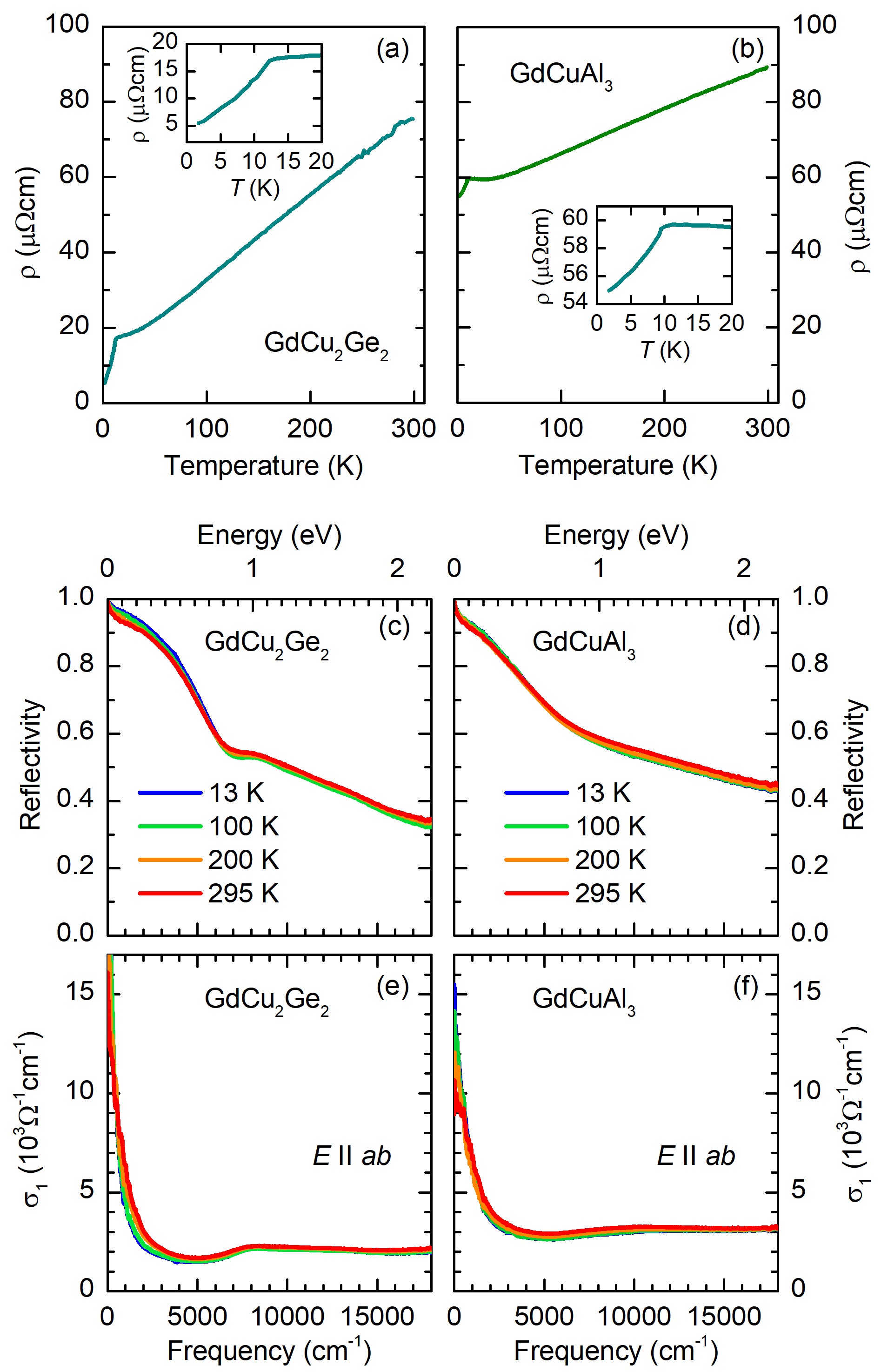}
	\caption{Dc resistivity curves of \GCG~(a) and \GCA~(b) measured in the $ab$-plane and normalized to the resistivity values obtained from our optical measurements via Hagen-Rubens fits. The kinks at low temperatures, highlighted in the insets, correspond to the antiferromagnetic transitions at $T_{\mathrm{N}} = 10$~K and 12~K of \GCG\ and \GCA, respectively. Panels (c) and (d) show the in-plane reflectivity of \GCG\ and \GCA\ at selected temperatures, while panels (e) and (f) present the corresponding calculated real part of the optical conductivity.}		
	\label{ROC}
\end{figure}

Fig.~\ref{ROC}(c) and (d) display the temperature-dependent reflectivity of \GCG\ and \GCA, respectively. At low frequencies, the high reflectivity values together with the Drude-like increase in the optical conductivities, presented in panels~(e) and (f), demonstrate the highly metallic nature of the samples. Conductivity values in the $\omega \rightarrow$~0 limit are obtained from the
Hagen-Rubens fit of the reflectivity and match well with our four-probe dc resistivity measurements performed on the same samples, as shown in the panels~(a) and (b) of Fig.~\ref{ROC}.

To gain further insight into the different electronic contributions to the spectra, we perform a Drude-Lorentz fit by simultaneously modeling the measured reflectivity and the complex optical response obtained from the Kramers-Kronig analysis.
\begin{equation}
\label{Eps}
\tilde{\varepsilon}(\omega)= \varepsilon_\infty - \frac{\omega^2_{p,{\rm Drude}}}{\omega^2 + i\omega/\tau_{\rm\, Drude}} + \sum\limits_j\frac{\Omega_j^2}{\omega_{0,j}^2 - \omega^2-i\omega\gamma_j}.
\end{equation}
Here, $\varepsilon_\infty$ accounts for the high-energy contributions to the real part of the dielectric permittivity [$\tilde{\varepsilon}=\varepsilon_1 + i\varepsilon_2$], and $\omega_{p,{\rm Drude}}$ and $1/\tau_{\rm\,Drude}$ are the plasma frequency and the scattering rate of the itinerant carriers, respectively. $\omega_{0,j}$, $\Omega_j$, and $\gamma_j$ describe the resonance frequency, width, and the strength of the $j^{th}$ excitation, respectively. The complex optical conductivity [$\tilde{\sigma}=\sigma_1 + i\sigma_2$] is then calculated as
\begin{equation}
\tilde{\sigma}(\omega)= -i\omega[\tilde{\varepsilon}-\varepsilon_\infty]/4\pi.
\label{Cond}
\end{equation}

\begin{figure}
	\centering
	\includegraphics[width=0.9\columnwidth]{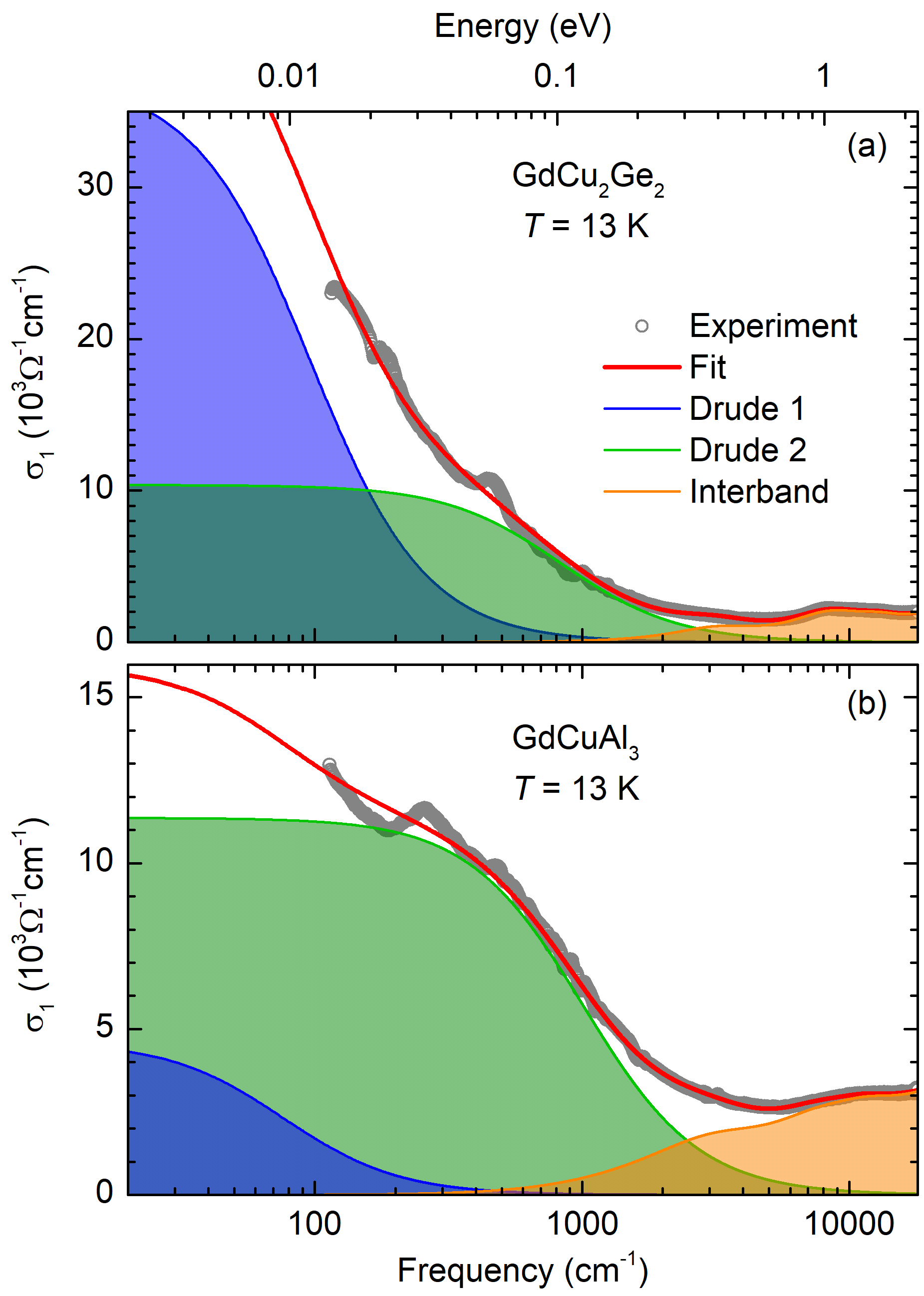}
	\caption{Optical conductivity of \GCG~(a) and \GCA~(b) at 13~K, modeled using two Drude contributions (blue and green) and five Lorentzians above 2000~\cm, describing the interband transitions (orange). }		
	\label{decomp}
\end{figure}

The decomposed experimental optical conductivity of \GCG\ and \GCA\ at 13~K is shown in Fig.~\ref{decomp}(a) and (b), respectively (see the decomposed optical spectra at other selected temperatures in the Supplemental Material \cite{SM}). At high frequencies ($\omega > 2000$~\cm), temperature-independent interband transitions are modeled by a total of five Lorentzians, shown in orange. At lower frequencies ($\omega < 2000$~\cm), the spectra exhibit a broad intraband contribution that, in the case of \GCG, cannot be adequately described by a single Drude term. Although less apparent within the measured spectral range of \GCA, a second Drude component (shown in blue in Fig.~\ref{decomp}) with significantly reduced spectral weight is required here as well to model the slight upturn below $\sim$200~\cm, as well as to reproduce the dc conductivity obtained from the four-contact transport measurements.

The occurrence of two Drude contributions with markedly different scattering rates is commonly observed in multiband systems, including topological materials such as Dirac and Weyl semimetals, as well as iron pnictides with the ThCr$_2$Si$_2$ structure~\cite{Schilling2017, Neubauer2018, Yang2024, Homes2020, Nakajima2014, Maulana2020, Barisic2010}. While the exact interpretation varies among different material classes, many of these systems host linearly dispersing bands that give rise to highly mobile carriers, whereas carriers associated with nonlinear bands tend to exhibit lower mobility. This picture is consistent with the electronic structure of the intermetallic compounds investigated here, as discussed below. At the same time, it should be noted that the two Drude components may also reflect two inherently different scattering channels rather than carriers originating from specific electronic bands~\cite{Kemmler2018, Maulana2020, Ni2020, Maslov2017}.

\begin{figure}
	\centering
	\includegraphics[width=1\columnwidth]{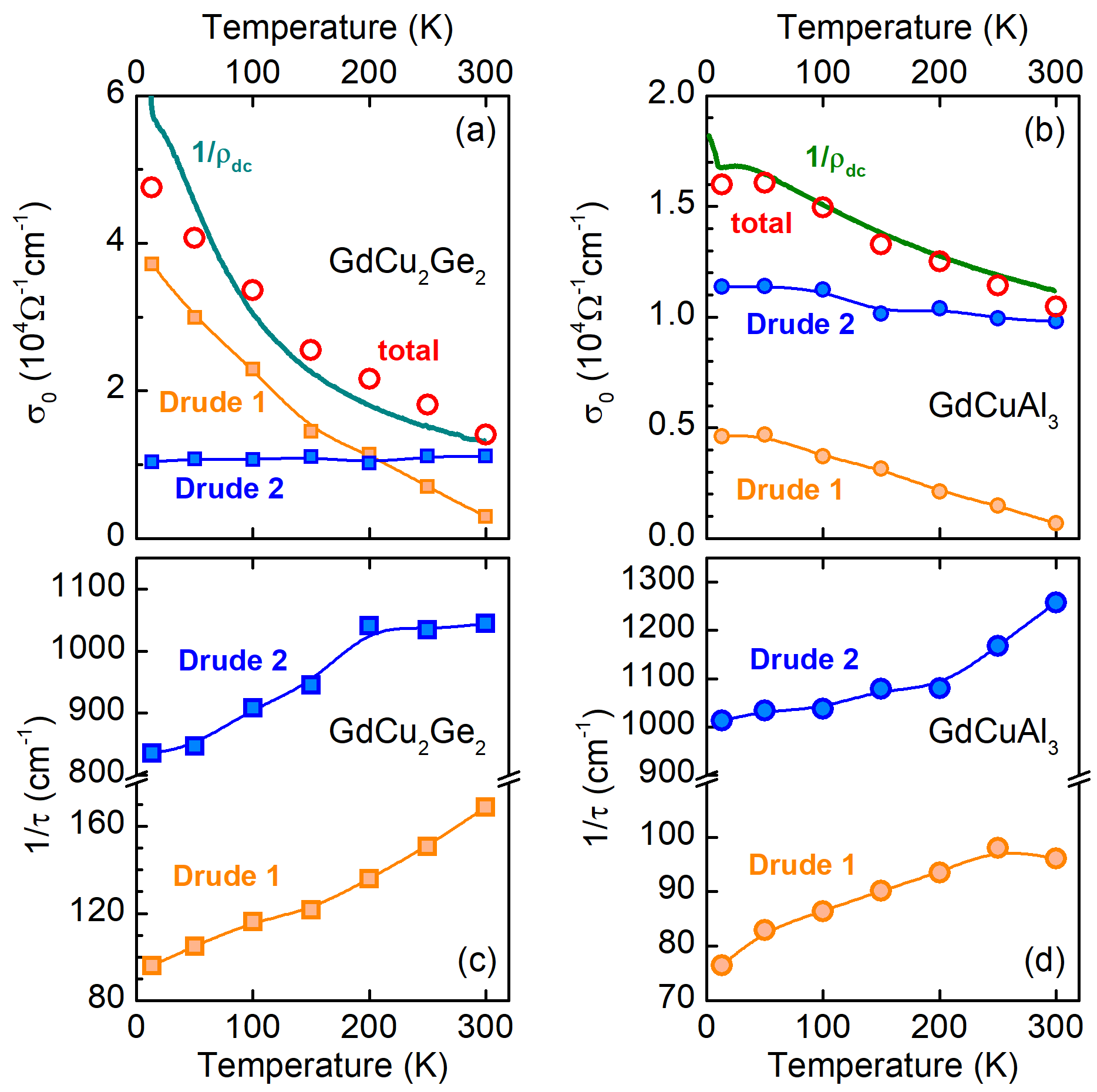}
	\caption{(a) and (b) Temperature-dependent dc conductivities of the Drude fits overlaid with the values obtained from the measured dc resistivity for \GCG\ and \GCA, respectively. The associated scattering rates are plotted in panels~(c) for \GCG\ and (d) for \GCA.} 	
	\label{Drudepara}
\end{figure}

\begin{figure}
	\centering
	\includegraphics[width=1\columnwidth]{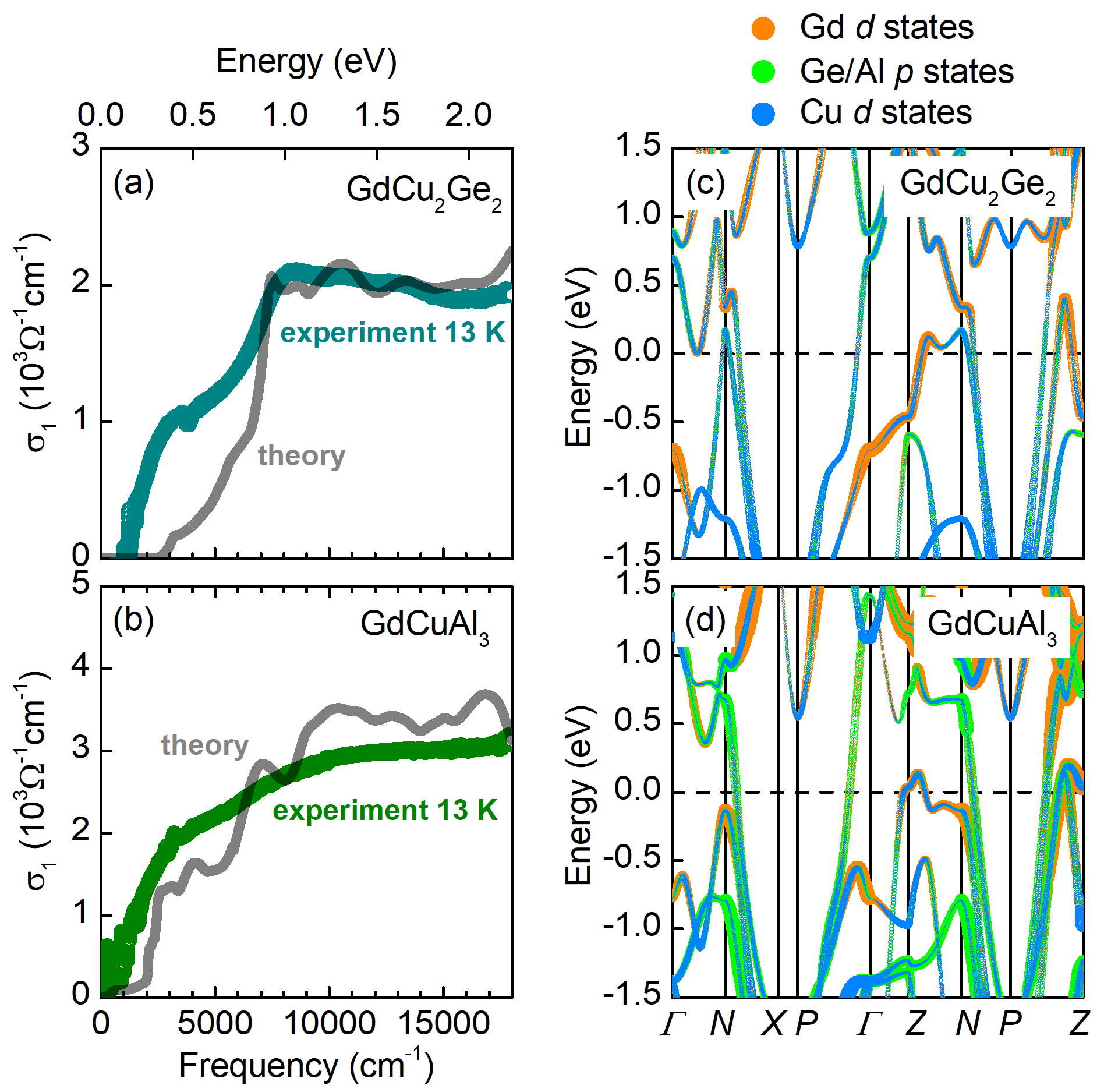}
	\caption{(a, b) Experimental interband optical transitions of \GCG\ and \GCA\ at 13~K (colored scatter), respectively, obtained by subtracting the Drude contributions form the spectra, overlaid with the DFT-calculated optical conductivities (grey solid line). Note that a broadening of up to 0.1~eV was applied to the calculated spectra above 5000~\cm\ to facilitate the direct comparison with the experimental data. Furthermore, the energy axis of the theoretical optical conductivity of \GCA\ is rescaled by a factor of 1.5. Band structures of \GCG\ and \GCA\ are given in panels~(c) and (d), respectively, with colored dots representing contributions from different atomic orbitals.}		
	\label{DFT}
\end{figure}

As summarized in Fig.~\ref{Drudepara}, the dc conductivity associated with the broad Drude term (Drude 2), remains temperature independent and exhibits similar scattering rates in both compounds, within the error bars of our fits. In contrast, the dc conductivity of the narrow Drude term (Drude 1) closely follows the temperature dependence of the measured resistivity and is significantly enhanced in \GCG, consistent with its lower resistivity.

The strength of the electronic correlations can be gauged by comparing the experimental and DFT-based plasma frequencies~\cite{Shao2020, Qazilbash2009, Wenzel2025}. Here the ratio $\omega_{\mathrm{p, exp}}^2/\omega_{\mathrm{p, DFT}}^2$ is close to 1 for uncorrelated systems, while it approaches 0 for Mott insulators, indicating the full suppression of itinerant charge carriers. Given the presence of two Drude contributions in our spectra, the total experimental plasma frequency is calculated by $\omega_{\mathrm{p, exp}} = \sqrt{\omega_{\mathrm{p, Drude 1}}^2 +  \omega_{\mathrm{p, Drude 2}}^2}$. Using the experimental and DFT-derived plasma frequencies listed in Tab.~\ref{tableII}, we calculate the ratio of 0.47 for both compounds at 13~K, indicating the presence of enhanced correlations. The determined ratios are highly robust with increasing temperatures, as demonstrated by the comparable values obtained at room temperature also found in Tab.~\ref{tableII}.

\begin{table}
\centering
\caption{Drude fit parameters at 13~K and 295~K together with the calculated plasma frequencies as well as the ratio $\omega_{\mathrm{p, exp}}^2/\omega_{\mathrm{p, DFT}}^2$ used as a gauge of electronic correlations.}
\label{tableII}
\begin{tabular*}{\columnwidth}{@{\extracolsep{\fill}}lcccc@{}}
\toprule
& \multicolumn{2}{c}{\GCG} & \multicolumn{2}{c}{\GCA} \\
& 13 K & 295 K & 13 K & 295 K \\
\midrule
$\sigma_{0,1}$ ($\Omega^{-1}$\cm) & 37162 & 2925 & 4619 & 679 \\
$\sigma_{0,2}$ ($\Omega^{-1}$\cm) & 10370 & 11121 & 11365 & 9800 \\
$\tau_1^{-1}$ (\cm) & 96 & 169 & 76 & 96 \\
$\tau_2^{-1}$ (\cm) & 836 & 1044 & 1013 & 1257 \\
$\omega_{\mathrm{p, Drude 1}}$ (\cm) & 14633 & 5446 & 4603 & 1979 \\
$\omega_{\mathrm{p, Drude 2}}$ (\cm) & 22808 & 26394 & 26284 & 27191 \\
$\omega_{\mathrm{p, exp}}$ (\cm) & 27098 & 26950 & 26684 & 27263 \\
$\omega_{\mathrm{p, DFT}}$ (\cm) & \multicolumn{2}{c}{39733} & \multicolumn{2}{c}{39059} \\
$\omega_{\mathrm{p, exp}}^2/\omega_{\mathrm{p, DFT}}^2$ & 0.47 & 0.46 & 0.47 & 0.49 \\
\bottomrule
\end{tabular*}

\end{table}

Further insight into the role of structural changes and differences in bonding on the low-energy electronic structure is obtained through a comparison between experimental and DFT-derived interband optical transitions summarized in Fig.~\ref{DFT}. Panels~(a) and (b) display the experimental interband optical conductivity of \GCG\ and \GCA, respectively, obtained by subtracting the modeled Drude contributions from the data. For \GCG, the interband absorption above 1~eV is well reproduced by the DFT calculations shown in grey in the same panel. At lower energies, however, the experimental data reveal a step-like onset whose spectral weight is strongly suppressed in the DFT-derived optical conductivity. As revealed by our band-resolved calculations in Fig.~S2~\cite{SM}, these low-energy interband excitations arise predominantly from transitions between linearly dispersing bands crossing the Fermi energy along $N$--$X$, $N$--$P$, and $P$--$Z$ (see Fig.~\ref{DFT} for the band structures).

For \GCA, the best agreement between the experimental and DFT-derived optical spectra is obtained by rescaling the calculated energies by a factor of 1.5, as shown in Fig.~\ref{DFT}(b). Such a rescaling is commonly observed in correlated materials and indicates a moderate renormalization of the band energies~\cite{Okamura2020, Stewart2011, Wenzel2024}. A similar
renormalization of band energies can also be envisaged for \GCG,
where the possible effect would primarily concern bands 'B' and 'C' in close proximity to the Fermi level (see Supplemental Material Fig.~S2  \cite{SM}), and would consequently shift the onset of interband transitions involving these two bands to lower energies.

In both compounds, the band structure is dominated by strongly hybridized Gd 5$d$, Cu 3$d$, and Al/Ge $p$ states, as demonstrated in Fig.~\ref{DFT}(c) and (d). The overall electronic structure exhibits a multi-band nature, comprising linearly dispersing bands and saddle points, and remains largely unchanged upon chemical substitution, highlighting its robustness against structural modifications. This renders these Gd-based intermetallics promising platforms for tuning electronic properties through controlled chemical substitution or carrier doping while preserving their characteristic low-energy band features. Notably, the electronic structure of \GCA\ exhibits substantially enhanced contributions from Al 3$p$ states, resulting in an upward shift of the Al 3$p$-derived bands. As shown in Fig.~S2, these shifts do not crucially alter the shape of the interband transitions across the measured energy range. Such tunability provides an avenue for exploring filling-dependent electronic and magnetic phases, including high-temperature superconductivity and topological spin textures such as skyrmions~\cite{Johrendt1997, Miranda2022, Dupe2016, Markiewicz1991, Irkhin2002}.

Furthermore, this upward shift of the 3$p$ states leads to the formation of band saddle points along $\Gamma$--$Z$--$N$ that lie in close proximity to the Fermi level. Together with the overall enhanced Al 3$p$ contribution to the Fermi surface, this may account for changes in scattering mechanism, resulting in the dampened Drude response and the overall higher resistivity observed in \GCA. Finally, the loss of centrosymmetry in \GCA\ gives rise to small spin-orbit-coupling-induced band splittings on the order of $\sim 20$~meV along certain $k$-directions (see Fig.~S3~\cite{SM}). These splittings do not produce additional interband transitions with sufficient spectral weight to be resolved in the optical spectra, consistent with the DFT-derived optical conductivity.

\section{Conclusions}
We studied the broadband in-plane optical conductivity of single crystals of two Gd-based ternary intermetallic compounds: the centrosymmetric \GCG\ and the non-centrosymmetric \GCA. The response of free carriers reveals two distinct scattering rates, which can be interpreted in terms of highly-damped and weakly-damped charge carriers, with the latter being the dominant contribution to the temperature dependence of the dc transport. The electronic structures of both compounds exhibit highly similar band dispersions, and enhanced electronic correlations, evidenced by reduced experimental plasma frequencies and band-energy renormalization. In \GCA, the augmented Al 3$p$ contribution modifies the orbital character near the Fermi surface, leading to band saddle points at the Fermi level and an overall upward shift of bands with strong Al 3$p$ contribution. The loss of centrosymmetry in \GCA, on the other hand, has only a minor effect on the overall electronic structure, producing spin-orbit-coupling-induced splittings of several tens of meV without introducing additional interband optical transitions. These results highlight the robustness of the electronic structure in ternary intermetallics against chemical substitution, making them promising platforms for future band-filling tuning and exploration of emergent electronic and magnetic phases. The coexistence of itinerant carriers with two distinct scattering rates motivates Hall-effect measurements to further disentangle their contributions to the transport properties.

\bibliography{Gdinter}

\begin{thebibliography}{73}%
\makeatletter
\providecommand \@ifxundefined [1]{%
 \@ifx{#1\undefined}
}%
\providecommand \@ifnum [1]{%
 \ifnum #1\expandafter \@firstoftwo
 \else \expandafter \@secondoftwo
 \fi
}%
\providecommand \@ifx [1]{%
 \ifx #1\expandafter \@firstoftwo
 \else \expandafter \@secondoftwo
 \fi
}%
\providecommand \natexlab [1]{#1}%
\providecommand \enquote  [1]{``#1''}%
\providecommand \bibnamefont  [1]{#1}%
\providecommand \bibfnamefont [1]{#1}%
\providecommand \citenamefont [1]{#1}%
\providecommand \href@noop [0]{\@secondoftwo}%
\providecommand \href [0]{\begingroup \@sanitize@url \@href}%
\providecommand \@href[1]{\@@startlink{#1}\@@href}%
\providecommand \@@href[1]{\endgroup#1\@@endlink}%
\providecommand \@sanitize@url [0]{\catcode `\\12\catcode `\$12\catcode
  `\&12\catcode `\#12\catcode `\^12\catcode `\_12\catcode `\%12\relax}%
\providecommand \@@startlink[1]{}%
\providecommand \@@endlink[0]{}%
\providecommand \url  [0]{\begingroup\@sanitize@url \@url }%
\providecommand \@url [1]{\endgroup\@href {#1}{\urlprefix }}%
\providecommand \urlprefix  [0]{URL }%
\providecommand \Eprint [0]{\href }%
\providecommand \doibase [0]{https://doi.org/}%
\providecommand \selectlanguage [0]{\@gobble}%
\providecommand \bibinfo  [0]{\@secondoftwo}%
\providecommand \bibfield  [0]{\@secondoftwo}%
\providecommand \translation [1]{[#1]}%
\providecommand \BibitemOpen [0]{}%
\providecommand \bibitemStop [0]{}%
\providecommand \bibitemNoStop [0]{.\EOS\space}%
\providecommand \EOS [0]{\spacefactor3000\relax}%
\providecommand \BibitemShut  [1]{\csname bibitem#1\endcsname}%
\let\auto@bib@innerbib\@empty
\bibitem [{\citenamefont {Lai}\ \emph {et~al.}(2022)\citenamefont {Lai},
  \citenamefont {Chan},\ and\ \citenamefont {Baumbach}}]{Lai2022}%
  \BibitemOpen
  \bibfield  {author} {\bibinfo {author} {\bibfnamefont {Y.}~\bibnamefont
  {Lai}}, \bibinfo {author} {\bibfnamefont {J.~Y.}\ \bibnamefont {Chan}},\ and\
  \bibinfo {author} {\bibfnamefont {R.~E.}\ \bibnamefont {Baumbach}},\
  }\bibfield  {title} {\bibinfo {title} {{Electronic landscape of the
  $f$-electron intermetallics with the ThCr$_2$Si$_2$ structure}},\ }\href
  {https://doi.org/10.1126/sciadv.abp8264} {\bibfield  {journal} {\bibinfo
  {journal} {Sci. Adv.}\ }\textbf {\bibinfo {volume} {8}},\ \bibinfo {pages}
  {eabp8264} (\bibinfo {year} {2022})}\BibitemShut {NoStop}%
\bibitem [{\citenamefont {Shatruk}(2019)}]{Shatruk2019}%
  \BibitemOpen
  \bibfield  {author} {\bibinfo {author} {\bibfnamefont {M.}~\bibnamefont
  {Shatruk}},\ }\bibfield  {title} {\bibinfo {title} {{ThCr$_2$Si$_2$ structure
  type: The “perovskite” of intermetallics}},\ }\href
  {https://doi.org/https://doi.org/10.1016/j.jssc.2019.02.012} {\bibfield
  {journal} {\bibinfo  {journal} {J. Solid State Chem.}\ }\textbf {\bibinfo
  {volume} {272}},\ \bibinfo {pages} {198} (\bibinfo {year}
  {2019})}\BibitemShut {NoStop}%
\bibitem [{\citenamefont {Johrendt}\ \emph {et~al.}(1997)\citenamefont
  {Johrendt}, \citenamefont {Felser}, \citenamefont {Jepsen}, \citenamefont
  {Andersen}, \citenamefont {Mewis},\ and\ \citenamefont
  {Rouxel}}]{Johrendt1997}%
  \BibitemOpen
  \bibfield  {author} {\bibinfo {author} {\bibfnamefont {D.}~\bibnamefont
  {Johrendt}}, \bibinfo {author} {\bibfnamefont {C.}~\bibnamefont {Felser}},
  \bibinfo {author} {\bibfnamefont {O.}~\bibnamefont {Jepsen}}, \bibinfo
  {author} {\bibfnamefont {O.~K.}\ \bibnamefont {Andersen}}, \bibinfo {author}
  {\bibfnamefont {A.}~\bibnamefont {Mewis}},\ and\ \bibinfo {author}
  {\bibfnamefont {J.}~\bibnamefont {Rouxel}},\ }\bibfield  {title} {\bibinfo
  {title} {{LMTO Band Structure Calculations of ThCr$_2$Si$_2$-Type Transition
  Metal Compounds}},\ }\href
  {https://doi.org/https://doi.org/10.1006/jssc.1997.7300} {\bibfield
  {journal} {\bibinfo  {journal} {J. Solid State Chem.}\ }\textbf {\bibinfo
  {volume} {130}},\ \bibinfo {pages} {254} (\bibinfo {year}
  {1997})}\BibitemShut {NoStop}%
\bibitem [{\citenamefont {Szytu{\l}a}\ and\ \citenamefont
  {Leciejewicz}(1989)}]{Szytula1989}%
  \BibitemOpen
  \bibfield  {author} {\bibinfo {author} {\bibfnamefont {A.}~\bibnamefont
  {Szytu{\l}a}}\ and\ \bibinfo {author} {\bibfnamefont {J.}~\bibnamefont
  {Leciejewicz}},\ }\bibfield  {title} {\bibinfo {title} {{Chapter 83 Magnetic
  properties of ternary intermetallic compounds of the RT$_2$X$_2$ type}}\
  }(\bibinfo  {publisher} {Elsevier},\ \bibinfo {year} {1989})\ pp.\ \bibinfo
  {pages} {133--211}\BibitemShut {NoStop}%
\bibitem [{\citenamefont {Tan}\ \emph {et~al.}(2018)\citenamefont {Tan},
  \citenamefont {Tener},\ and\ \citenamefont {Shatruk}}]{Tan2018}%
  \BibitemOpen
  \bibfield  {author} {\bibinfo {author} {\bibfnamefont {X.}~\bibnamefont
  {Tan}}, \bibinfo {author} {\bibfnamefont {Z.~P.}\ \bibnamefont {Tener}},\
  and\ \bibinfo {author} {\bibfnamefont {M.}~\bibnamefont {Shatruk}},\
  }\bibfield  {title} {\bibinfo {title} {{Correlating Itinerant Magnetism in
  RCo$_2$Pn$_2$ Pnictides (R = La, Ce, Pr, Nd, Eu, Ca; Pn = P, As) to Their
  Crystal and Electronic Structures}},\ }\href
  {https://doi.org/10.1021/acs.accounts.7b00533} {\bibfield  {journal}
  {\bibinfo  {journal} {Acc. Chem. Res.}\ }\textbf {\bibinfo {volume} {51}},\
  \bibinfo {pages} {230} (\bibinfo {year} {2018})}\BibitemShut {NoStop}%
\bibitem [{\citenamefont {Petit}\ \emph {et~al.}(2015)\citenamefont {Petit},
  \citenamefont {Paudyal}, \citenamefont {Mudryk}, \citenamefont {Gschneidner},
  \citenamefont {Pecharsky}, \citenamefont {L\"uders}, \citenamefont {Szotek},
  \citenamefont {Banerjee},\ and\ \citenamefont {Staunton}}]{Petit2015}%
  \BibitemOpen
  \bibfield  {author} {\bibinfo {author} {\bibfnamefont {L.}~\bibnamefont
  {Petit}}, \bibinfo {author} {\bibfnamefont {D.}~\bibnamefont {Paudyal}},
  \bibinfo {author} {\bibfnamefont {Y.}~\bibnamefont {Mudryk}}, \bibinfo
  {author} {\bibfnamefont {K.~A.}\ \bibnamefont {Gschneidner}}, \bibinfo
  {author} {\bibfnamefont {V.~K.}\ \bibnamefont {Pecharsky}}, \bibinfo {author}
  {\bibfnamefont {M.}~\bibnamefont {L\"uders}}, \bibinfo {author}
  {\bibfnamefont {Z.}~\bibnamefont {Szotek}}, \bibinfo {author} {\bibfnamefont
  {R.}~\bibnamefont {Banerjee}},\ and\ \bibinfo {author} {\bibfnamefont
  {J.~B.}\ \bibnamefont {Staunton}},\ }\bibfield  {title} {\bibinfo {title}
  {{Complex Magnetism of Lanthanide Intermetallics and the Role of their
  Valence Electrons: Ab Initio Theory and Experiment}},\ }\href
  {https://doi.org/10.1103/PhysRevLett.115.207201} {\bibfield  {journal}
  {\bibinfo  {journal} {Phys. Rev. Lett.}\ }\textbf {\bibinfo {volume} {115}},\
  \bibinfo {pages} {207201} (\bibinfo {year} {2015})}\BibitemShut {NoStop}%
\bibitem [{\citenamefont {Pfleiderer}(2009)}]{Pfleiderer2009}%
  \BibitemOpen
  \bibfield  {author} {\bibinfo {author} {\bibfnamefont {C.}~\bibnamefont
  {Pfleiderer}},\ }\bibfield  {title} {\bibinfo {title} {{Superconducting
  phases of $f$-electron compounds}},\ }\href
  {https://doi.org/10.1103/RevModPhys.81.1551} {\bibfield  {journal} {\bibinfo
  {journal} {Rev. Mod. Phys.}\ }\textbf {\bibinfo {volume} {81}},\ \bibinfo
  {pages} {1551} (\bibinfo {year} {2009})}\BibitemShut {NoStop}%
\bibitem [{\citenamefont {Dzero}\ \emph {et~al.}(2010)\citenamefont {Dzero},
  \citenamefont {Sun}, \citenamefont {Galitski},\ and\ \citenamefont
  {Coleman}}]{Dzero2010}%
  \BibitemOpen
  \bibfield  {author} {\bibinfo {author} {\bibfnamefont {M.}~\bibnamefont
  {Dzero}}, \bibinfo {author} {\bibfnamefont {K.}~\bibnamefont {Sun}}, \bibinfo
  {author} {\bibfnamefont {V.}~\bibnamefont {Galitski}},\ and\ \bibinfo
  {author} {\bibfnamefont {P.}~\bibnamefont {Coleman}},\ }\bibfield  {title}
  {\bibinfo {title} {{Topological Kondo Insulators}},\ }\href
  {https://doi.org/10.1103/PhysRevLett.104.106408} {\bibfield  {journal}
  {\bibinfo  {journal} {Phys. Rev. Lett.}\ }\textbf {\bibinfo {volume} {104}},\
  \bibinfo {pages} {106408} (\bibinfo {year} {2010})}\BibitemShut {NoStop}%
\bibitem [{\citenamefont {Puphal}\ \emph {et~al.}(2020)\citenamefont {Puphal},
  \citenamefont {Pomjakushin}, \citenamefont {Kanazawa}, \citenamefont
  {Ukleev}, \citenamefont {Gawryluk}, \citenamefont {Ma}, \citenamefont
  {Naamneh}, \citenamefont {Plumb}, \citenamefont {Keller}, \citenamefont
  {Cubitt}, \citenamefont {Pomjakushina},\ and\ \citenamefont
  {White}}]{Puphal2020}%
  \BibitemOpen
  \bibfield  {author} {\bibinfo {author} {\bibfnamefont {P.}~\bibnamefont
  {Puphal}}, \bibinfo {author} {\bibfnamefont {V.}~\bibnamefont {Pomjakushin}},
  \bibinfo {author} {\bibfnamefont {N.}~\bibnamefont {Kanazawa}}, \bibinfo
  {author} {\bibfnamefont {V.}~\bibnamefont {Ukleev}}, \bibinfo {author}
  {\bibfnamefont {D.~J.}\ \bibnamefont {Gawryluk}}, \bibinfo {author}
  {\bibfnamefont {J.}~\bibnamefont {Ma}}, \bibinfo {author} {\bibfnamefont
  {M.}~\bibnamefont {Naamneh}}, \bibinfo {author} {\bibfnamefont {N.~C.}\
  \bibnamefont {Plumb}}, \bibinfo {author} {\bibfnamefont {L.}~\bibnamefont
  {Keller}}, \bibinfo {author} {\bibfnamefont {R.}~\bibnamefont {Cubitt}},
  \bibinfo {author} {\bibfnamefont {E.}~\bibnamefont {Pomjakushina}},\ and\
  \bibinfo {author} {\bibfnamefont {J.~S.}\ \bibnamefont {White}},\ }\bibfield
  {title} {\bibinfo {title} {{Topological Magnetic Phase in the Candidate Weyl
  Semimetal CeAlGe}},\ }\href {https://doi.org/10.1103/PhysRevLett.124.017202}
  {\bibfield  {journal} {\bibinfo  {journal} {Phys. Rev. Lett.}\ }\textbf
  {\bibinfo {volume} {124}},\ \bibinfo {pages} {017202} (\bibinfo {year}
  {2020})}\BibitemShut {NoStop}%
\bibitem [{\citenamefont {H{\"a}ussermann}\ \emph {et~al.}(2002)\citenamefont
  {H{\"a}ussermann}, \citenamefont {Amerioun}, \citenamefont {Eriksson},
  \citenamefont {Lee},\ and\ \citenamefont {Miller}}]{Haussermann2002}%
  \BibitemOpen
  \bibfield  {author} {\bibinfo {author} {\bibfnamefont {U.}~\bibnamefont
  {H{\"a}ussermann}}, \bibinfo {author} {\bibfnamefont {S.}~\bibnamefont
  {Amerioun}}, \bibinfo {author} {\bibfnamefont {L.}~\bibnamefont {Eriksson}},
  \bibinfo {author} {\bibfnamefont {C.-S.}\ \bibnamefont {Lee}},\ and\ \bibinfo
  {author} {\bibfnamefont {G.~J.}\ \bibnamefont {Miller}},\ }\bibfield  {title}
  {\bibinfo {title} {{The s-p Bonded Representatives of the Prominent BaAl$_4$
  Structure Type: A Case Study on Structural Stability of Polar Intermetallic
  Network Structures}},\ }\href {https://doi.org/10.1021/ja012392v} {\bibfield
  {journal} {\bibinfo  {journal} {J. Am. Chem. Soc.}\ }\textbf {\bibinfo
  {volume} {124}},\ \bibinfo {pages} {4371} (\bibinfo {year}
  {2002})}\BibitemShut {NoStop}%
\bibitem [{\citenamefont {Parth{\'{e}}}\ \emph {et~al.}(1983)\citenamefont
  {Parth{\'{e}}}, \citenamefont {Chabot}, \citenamefont {Braun},\ and\
  \citenamefont {Engel}}]{Parthe1983}%
  \BibitemOpen
  \bibfield  {author} {\bibinfo {author} {\bibfnamefont {E.}~\bibnamefont
  {Parth{\'{e}}}}, \bibinfo {author} {\bibfnamefont {B.}~\bibnamefont
  {Chabot}}, \bibinfo {author} {\bibfnamefont {H.~F.}\ \bibnamefont {Braun}},\
  and\ \bibinfo {author} {\bibfnamefont {N.}~\bibnamefont {Engel}},\ }\bibfield
   {title} {\bibinfo {title} {{Ternary BaAl${\sb 4}$-type derivative
  structures}},\ }\href {https://doi.org/10.1107/S010876818300302X} {\bibfield
  {journal} {\bibinfo  {journal} {Acta Cryst. B}\ }\textbf {\bibinfo {volume}
  {39}},\ \bibinfo {pages} {588} (\bibinfo {year} {1983})}\BibitemShut
  {NoStop}%
\bibitem [{\citenamefont {Khim}\ \emph {et~al.}(2021)\citenamefont {Khim},
  \citenamefont {Landaeta}, \citenamefont {Banda}, \citenamefont {Bannor},
  \citenamefont {Brando}, \citenamefont {Brydon}, \citenamefont {Hafner},
  \citenamefont {Küchler}, \citenamefont {Cardoso-Gil}, \citenamefont
  {Stockert}, \citenamefont {Mackenzie}, \citenamefont {Agterberg},
  \citenamefont {Geibel},\ and\ \citenamefont {Hassinger}}]{Khim2021}%
  \BibitemOpen
  \bibfield  {author} {\bibinfo {author} {\bibfnamefont {S.}~\bibnamefont
  {Khim}}, \bibinfo {author} {\bibfnamefont {J.~F.}\ \bibnamefont {Landaeta}},
  \bibinfo {author} {\bibfnamefont {J.}~\bibnamefont {Banda}}, \bibinfo
  {author} {\bibfnamefont {N.}~\bibnamefont {Bannor}}, \bibinfo {author}
  {\bibfnamefont {M.}~\bibnamefont {Brando}}, \bibinfo {author} {\bibfnamefont
  {P.~M.~R.}\ \bibnamefont {Brydon}}, \bibinfo {author} {\bibfnamefont
  {D.}~\bibnamefont {Hafner}}, \bibinfo {author} {\bibfnamefont
  {R.}~\bibnamefont {Küchler}}, \bibinfo {author} {\bibfnamefont
  {R.}~\bibnamefont {Cardoso-Gil}}, \bibinfo {author} {\bibfnamefont
  {U.}~\bibnamefont {Stockert}}, \bibinfo {author} {\bibfnamefont {A.~P.}\
  \bibnamefont {Mackenzie}}, \bibinfo {author} {\bibfnamefont {D.~F.}\
  \bibnamefont {Agterberg}}, \bibinfo {author} {\bibfnamefont {C.}~\bibnamefont
  {Geibel}},\ and\ \bibinfo {author} {\bibfnamefont {E.}~\bibnamefont
  {Hassinger}},\ }\bibfield  {title} {\bibinfo {title} {{Field-induced
  transition within the superconducting state of CeRh$_2$As$_2$}},\ }\href
  {https://doi.org/10.1126/science.abe7518} {\bibfield  {journal} {\bibinfo
  {journal} {Science}\ }\textbf {\bibinfo {volume} {373}},\ \bibinfo {pages}
  {1012} (\bibinfo {year} {2021})}\BibitemShut {NoStop}%
\bibitem [{\citenamefont {Nogaki}\ \emph {et~al.}(2021)\citenamefont {Nogaki},
  \citenamefont {Daido}, \citenamefont {Ishizuka},\ and\ \citenamefont
  {Yanase}}]{Nogaki2021}%
  \BibitemOpen
  \bibfield  {author} {\bibinfo {author} {\bibfnamefont {K.}~\bibnamefont
  {Nogaki}}, \bibinfo {author} {\bibfnamefont {A.}~\bibnamefont {Daido}},
  \bibinfo {author} {\bibfnamefont {J.}~\bibnamefont {Ishizuka}},\ and\
  \bibinfo {author} {\bibfnamefont {Y.}~\bibnamefont {Yanase}},\ }\bibfield
  {title} {\bibinfo {title} {{Topological crystalline superconductivity in
  locally noncentrosymmetric ${\mathrm{CeRh}}_{2}{\mathrm{As}}_{2}$}},\ }\href
  {https://doi.org/10.1103/PhysRevResearch.3.L032071} {\bibfield  {journal}
  {\bibinfo  {journal} {Phys. Rev. Res.}\ }\textbf {\bibinfo {volume} {3}},\
  \bibinfo {pages} {L032071} (\bibinfo {year} {2021})}\BibitemShut {NoStop}%
\bibitem [{\citenamefont {Yuan}\ \emph {et~al.}(2003)\citenamefont {Yuan},
  \citenamefont {Grosche}, \citenamefont {Deppe}, \citenamefont {Geibel},
  \citenamefont {Sparn},\ and\ \citenamefont {Steglich}}]{Yuan2003}%
  \BibitemOpen
  \bibfield  {author} {\bibinfo {author} {\bibfnamefont {H.~Q.}\ \bibnamefont
  {Yuan}}, \bibinfo {author} {\bibfnamefont {F.~M.}\ \bibnamefont {Grosche}},
  \bibinfo {author} {\bibfnamefont {M.}~\bibnamefont {Deppe}}, \bibinfo
  {author} {\bibfnamefont {C.}~\bibnamefont {Geibel}}, \bibinfo {author}
  {\bibfnamefont {G.}~\bibnamefont {Sparn}},\ and\ \bibinfo {author}
  {\bibfnamefont {F.}~\bibnamefont {Steglich}},\ }\bibfield  {title} {\bibinfo
  {title} {{Observation of Two Distinct Superconducting Phases in
  CeCu$_2$Si$_2$}},\ }\href {https://doi.org/10.1126/science.1091648}
  {\bibfield  {journal} {\bibinfo  {journal} {Science}\ }\textbf {\bibinfo
  {volume} {302}},\ \bibinfo {pages} {2104} (\bibinfo {year}
  {2003})}\BibitemShut {NoStop}%
\bibitem [{\citenamefont {Schuberth}\ \emph {et~al.}(2016)\citenamefont
  {Schuberth}, \citenamefont {Tippmann}, \citenamefont {Steinke}, \citenamefont
  {Lausberg}, \citenamefont {Steppke}, \citenamefont {Brando}, \citenamefont
  {Krellner}, \citenamefont {Geibel}, \citenamefont {Yu}, \citenamefont {Si},\
  and\ \citenamefont {Steglich}}]{Schuberth2016}%
  \BibitemOpen
  \bibfield  {author} {\bibinfo {author} {\bibfnamefont {E.}~\bibnamefont
  {Schuberth}}, \bibinfo {author} {\bibfnamefont {M.}~\bibnamefont {Tippmann}},
  \bibinfo {author} {\bibfnamefont {L.}~\bibnamefont {Steinke}}, \bibinfo
  {author} {\bibfnamefont {S.}~\bibnamefont {Lausberg}}, \bibinfo {author}
  {\bibfnamefont {A.}~\bibnamefont {Steppke}}, \bibinfo {author} {\bibfnamefont
  {M.}~\bibnamefont {Brando}}, \bibinfo {author} {\bibfnamefont
  {C.}~\bibnamefont {Krellner}}, \bibinfo {author} {\bibfnamefont
  {C.}~\bibnamefont {Geibel}}, \bibinfo {author} {\bibfnamefont
  {R.}~\bibnamefont {Yu}}, \bibinfo {author} {\bibfnamefont {Q.}~\bibnamefont
  {Si}},\ and\ \bibinfo {author} {\bibfnamefont {F.}~\bibnamefont {Steglich}},\
  }\bibfield  {title} {\bibinfo {title} {{Emergence of superconductivity in the
  canonical heavy-electron metal YbRh$_2$Si$_2$}},\ }\href
  {https://doi.org/10.1126/science.aaa9733} {\bibfield  {journal} {\bibinfo
  {journal} {Science}\ }\textbf {\bibinfo {volume} {351}},\ \bibinfo {pages}
  {485} (\bibinfo {year} {2016})}\BibitemShut {NoStop}%
\bibitem [{\citenamefont {Steglich}(2023)}]{Steglich2023}%
  \BibitemOpen
  \bibfield  {author} {\bibinfo {author} {\bibfnamefont {F.}~\bibnamefont
  {Steglich}},\ }\bibfield  {title} {\bibinfo {title} {{Unconventional
  Superconductivity in the Kondo-lattice System CeCu$_2$Si$_2$ - a Personal
  Perspective}},\ }\href {https://doi.org/10.3938/NPSM.73.1067} {\bibfield
  {journal} {\bibinfo  {journal} {New Phys.: Sae Mulli}\ }\textbf {\bibinfo
  {volume} {73}},\ \bibinfo {pages} {1067} (\bibinfo {year}
  {2023})}\BibitemShut {NoStop}%
\bibitem [{\citenamefont {White}\ \emph {et~al.}(2015)\citenamefont {White},
  \citenamefont {Thompson},\ and\ \citenamefont {Maple}}]{White2015}%
  \BibitemOpen
  \bibfield  {author} {\bibinfo {author} {\bibfnamefont {B.}~\bibnamefont
  {White}}, \bibinfo {author} {\bibfnamefont {J.}~\bibnamefont {Thompson}},\
  and\ \bibinfo {author} {\bibfnamefont {M.}~\bibnamefont {Maple}},\ }\bibfield
   {title} {\bibinfo {title} {{Unconventional superconductivity in
  heavy-fermion compounds}},\ }\href
  {https://doi.org/https://doi.org/10.1016/j.physc.2015.02.044} {\bibfield
  {journal} {\bibinfo  {journal} {Phys. C}\ }\textbf {\bibinfo {volume}
  {514}},\ \bibinfo {pages} {246} (\bibinfo {year} {2015})}\BibitemShut
  {NoStop}%
\bibitem [{\citenamefont {Kimura}\ and\ \citenamefont
  {Bonalde}(2012)}]{Kimura2012}%
  \BibitemOpen
  \bibfield  {author} {\bibinfo {author} {\bibfnamefont {N.}~\bibnamefont
  {Kimura}}\ and\ \bibinfo {author} {\bibfnamefont {I.}~\bibnamefont
  {Bonalde}},\ }\bibinfo {title} {Non-centrosymmetric heavy-fermion
  superconductors},\ in\ \href {https://doi.org/10.1007/978-3-642-24624-1_2}
  {\emph {\bibinfo {booktitle} {Non-Centrosymmetric Superconductors:
  Introduction and Overview}}},\ \bibinfo {editor} {edited by\ \bibinfo
  {editor} {\bibfnamefont {E.}~\bibnamefont {Bauer}}\ and\ \bibinfo {editor}
  {\bibfnamefont {M.}~\bibnamefont {Sigrist}}}\ (\bibinfo  {publisher}
  {Springer Berlin Heidelberg},\ \bibinfo {address} {Berlin, Heidelberg},\
  \bibinfo {year} {2012})\ pp.\ \bibinfo {pages} {35--79}\BibitemShut {NoStop}%
\bibitem [{\citenamefont {Anand}\ \emph {et~al.}(2017)\citenamefont {Anand},
  \citenamefont {Adroja}, \citenamefont {Bhattacharyya}, \citenamefont
  {Klemke},\ and\ \citenamefont {Lake}}]{Anand2017}%
  \BibitemOpen
  \bibfield  {author} {\bibinfo {author} {\bibfnamefont {V.~K.}\ \bibnamefont
  {Anand}}, \bibinfo {author} {\bibfnamefont {D.~T.}\ \bibnamefont {Adroja}},
  \bibinfo {author} {\bibfnamefont {A.}~\bibnamefont {Bhattacharyya}}, \bibinfo
  {author} {\bibfnamefont {B.}~\bibnamefont {Klemke}},\ and\ \bibinfo {author}
  {\bibfnamefont {B.}~\bibnamefont {Lake}},\ }\bibfield  {title} {\bibinfo
  {title} {{Kondo lattice heavy fermion behavior in CeRh$_2$Ga$_2$}},\ }\href
  {https://doi.org/10.1088/1361-648X/aa5b5d} {\bibfield  {journal} {\bibinfo
  {journal} {J. Phys. Condens. Matter}\ }\textbf {\bibinfo {volume} {29}},\
  \bibinfo {pages} {135601} (\bibinfo {year} {2017})}\BibitemShut {NoStop}%
\bibitem [{\citenamefont {Lef\'evre}\ and\ \citenamefont {von
  Rohr}(2022)}]{Lefevre2022}%
  \BibitemOpen
  \bibfield  {author} {\bibinfo {author} {\bibfnamefont {R.}~\bibnamefont
  {Lef\'evre}}\ and\ \bibinfo {author} {\bibfnamefont {F.~O.}\ \bibnamefont
  {von Rohr}},\ }\bibfield  {title} {\bibinfo {title} {{A Heavy Fermion
  Zn-Deficient CaBe$_2$Ge$_2$-Type Phase with Rare Ce-Based Ferromagnetism and
  Large Magnetoresistance}},\ }\href
  {https://doi.org/10.1021/acs.chemmater.1c04328} {\bibfield  {journal}
  {\bibinfo  {journal} {Chem. Mater.}\ }\textbf {\bibinfo {volume} {34}},\
  \bibinfo {pages} {2352} (\bibinfo {year} {2022})}\BibitemShut {NoStop}%
\bibitem [{\citenamefont {Gupta}\ \emph {et~al.}(1983)\citenamefont {Gupta},
  \citenamefont {MacLaughlin}, \citenamefont {Tien}, \citenamefont {Godart},
  \citenamefont {Edwards},\ and\ \citenamefont {Parks}}]{Gupta1983}%
  \BibitemOpen
  \bibfield  {author} {\bibinfo {author} {\bibfnamefont {L.~C.}\ \bibnamefont
  {Gupta}}, \bibinfo {author} {\bibfnamefont {D.~E.}\ \bibnamefont
  {MacLaughlin}}, \bibinfo {author} {\bibfnamefont {C.}~\bibnamefont {Tien}},
  \bibinfo {author} {\bibfnamefont {C.}~\bibnamefont {Godart}}, \bibinfo
  {author} {\bibfnamefont {M.~A.}\ \bibnamefont {Edwards}},\ and\ \bibinfo
  {author} {\bibfnamefont {R.~D.}\ \bibnamefont {Parks}},\ }\bibfield  {title}
  {\bibinfo {title} {{Magnetic behavior of the Kondo-lattice system
  Ce${\mathrm{Ru}}_{2}$${\mathrm{Si}}_{2}$}},\ }\href
  {https://doi.org/10.1103/PhysRevB.28.3673} {\bibfield  {journal} {\bibinfo
  {journal} {Phys. Rev. B}\ }\textbf {\bibinfo {volume} {28}},\ \bibinfo
  {pages} {3673} (\bibinfo {year} {1983})}\BibitemShut {NoStop}%
\bibitem [{\citenamefont {Knebel}\ \emph {et~al.}(1999)\citenamefont {Knebel},
  \citenamefont {Brando}, \citenamefont {Hemberger}, \citenamefont {Nicklas},
  \citenamefont {Trinkl},\ and\ \citenamefont {Loidl}}]{Knebel1999}%
  \BibitemOpen
  \bibfield  {author} {\bibinfo {author} {\bibfnamefont {G.}~\bibnamefont
  {Knebel}}, \bibinfo {author} {\bibfnamefont {M.}~\bibnamefont {Brando}},
  \bibinfo {author} {\bibfnamefont {J.}~\bibnamefont {Hemberger}}, \bibinfo
  {author} {\bibfnamefont {M.}~\bibnamefont {Nicklas}}, \bibinfo {author}
  {\bibfnamefont {W.}~\bibnamefont {Trinkl}},\ and\ \bibinfo {author}
  {\bibfnamefont {A.}~\bibnamefont {Loidl}},\ }\bibfield  {title} {\bibinfo
  {title} {{Magnetic, calorimetric, and transport properties of
  ${\mathrm{C}\mathrm{e}(\mathrm{P}\mathrm{d}}_{1\ensuremath{-}x}{\mathrm{Ni}}_{x}{)}_{2}{\mathrm{Ge}}_{2}$
  and
  ${\mathrm{CeNi}}_{2}{(\mathrm{G}\mathrm{e}}_{1\ensuremath{-}y}{\mathrm{Si}}_{y}{)}_{2}$}},\
  }\href {https://doi.org/10.1103/PhysRevB.59.12390} {\bibfield  {journal}
  {\bibinfo  {journal} {Phys. Rev. B}\ }\textbf {\bibinfo {volume} {59}},\
  \bibinfo {pages} {12390} (\bibinfo {year} {1999})}\BibitemShut {NoStop}%
\bibitem [{\citenamefont {Mihalik}\ \emph {et~al.}(2009)\citenamefont
  {Mihalik}, \citenamefont {Diviš},\ and\ \citenamefont
  {Sechovský}}]{Mihalik2009}%
  \BibitemOpen
  \bibfield  {author} {\bibinfo {author} {\bibfnamefont {M.}~\bibnamefont
  {Mihalik}}, \bibinfo {author} {\bibfnamefont {M.}~\bibnamefont {Diviš}},\
  and\ \bibinfo {author} {\bibfnamefont {V.}~\bibnamefont {Sechovský}},\
  }\bibfield  {title} {\bibinfo {title} {{Electronic and crystal structure of
  $\alpha$- and $\beta$-CeIr$_2$Si$_2$}},\ }\href
  {https://doi.org/https://doi.org/10.1016/j.physb.2009.07.052} {\bibfield
  {journal} {\bibinfo  {journal} {Phys. B}\ }\textbf {\bibinfo {volume}
  {404}},\ \bibinfo {pages} {3191} (\bibinfo {year} {2009})}\BibitemShut
  {NoStop}%
\bibitem [{\citenamefont {Gegenwart}\ \emph {et~al.}(1998)\citenamefont
  {Gegenwart}, \citenamefont {Langhammer}, \citenamefont {Geibel},
  \citenamefont {Helfrich}, \citenamefont {Lang}, \citenamefont {Sparn},
  \citenamefont {Steglich}, \citenamefont {Horn}, \citenamefont {Donnevert},
  \citenamefont {Link},\ and\ \citenamefont {Assmus}}]{Gegenwart1998}%
  \BibitemOpen
  \bibfield  {author} {\bibinfo {author} {\bibfnamefont {P.}~\bibnamefont
  {Gegenwart}}, \bibinfo {author} {\bibfnamefont {C.}~\bibnamefont
  {Langhammer}}, \bibinfo {author} {\bibfnamefont {C.}~\bibnamefont {Geibel}},
  \bibinfo {author} {\bibfnamefont {R.}~\bibnamefont {Helfrich}}, \bibinfo
  {author} {\bibfnamefont {M.}~\bibnamefont {Lang}}, \bibinfo {author}
  {\bibfnamefont {G.}~\bibnamefont {Sparn}}, \bibinfo {author} {\bibfnamefont
  {F.}~\bibnamefont {Steglich}}, \bibinfo {author} {\bibfnamefont
  {R.}~\bibnamefont {Horn}}, \bibinfo {author} {\bibfnamefont {L.}~\bibnamefont
  {Donnevert}}, \bibinfo {author} {\bibfnamefont {A.}~\bibnamefont {Link}},\
  and\ \bibinfo {author} {\bibfnamefont {W.}~\bibnamefont {Assmus}},\
  }\bibfield  {title} {\bibinfo {title} {{Breakup of Heavy Fermions on the
  Brink of ``Phase $\mathit{A}$'' in $\mathrm{CeCu}{}_{2}\mathrm{Si}{}_{2}$}},\
  }\href {https://doi.org/10.1103/PhysRevLett.81.1501} {\bibfield  {journal}
  {\bibinfo  {journal} {Phys. Rev. Lett.}\ }\textbf {\bibinfo {volume} {81}},\
  \bibinfo {pages} {1501} (\bibinfo {year} {1998})}\BibitemShut {NoStop}%
\bibitem [{\citenamefont {Lee}\ \emph {et~al.}(2020)\citenamefont {Lee},
  \citenamefont {Proke\ifmmode~\check{s}\else \v{s}\fi{}}, \citenamefont
  {Park}, \citenamefont {Zaliznyak}, \citenamefont {Dissanayake}, \citenamefont
  {Matsuda}, \citenamefont {Frontzek}, \citenamefont {Stoupin}, \citenamefont
  {Chappell}, \citenamefont {Baumbach}, \citenamefont {Park}, \citenamefont
  {Mydosh}, \citenamefont {Granroth},\ and\ \citenamefont {Ruff}}]{Lee2020}%
  \BibitemOpen
  \bibfield  {author} {\bibinfo {author} {\bibfnamefont {J.}~\bibnamefont
  {Lee}}, \bibinfo {author} {\bibfnamefont {K.}~\bibnamefont
  {Proke\ifmmode~\check{s}\else \v{s}\fi{}}}, \bibinfo {author} {\bibfnamefont
  {S.}~\bibnamefont {Park}}, \bibinfo {author} {\bibfnamefont {I.}~\bibnamefont
  {Zaliznyak}}, \bibinfo {author} {\bibfnamefont {S.}~\bibnamefont
  {Dissanayake}}, \bibinfo {author} {\bibfnamefont {M.}~\bibnamefont
  {Matsuda}}, \bibinfo {author} {\bibfnamefont {M.}~\bibnamefont {Frontzek}},
  \bibinfo {author} {\bibfnamefont {S.}~\bibnamefont {Stoupin}}, \bibinfo
  {author} {\bibfnamefont {G.~L.}\ \bibnamefont {Chappell}}, \bibinfo {author}
  {\bibfnamefont {R.~E.}\ \bibnamefont {Baumbach}}, \bibinfo {author}
  {\bibfnamefont {C.}~\bibnamefont {Park}}, \bibinfo {author} {\bibfnamefont
  {J.~A.}\ \bibnamefont {Mydosh}}, \bibinfo {author} {\bibfnamefont {G.~E.}\
  \bibnamefont {Granroth}},\ and\ \bibinfo {author} {\bibfnamefont {J.~P.~C.}\
  \bibnamefont {Ruff}},\ }\bibfield  {title} {\bibinfo {title} {{Charge density
  wave with anomalous temperature dependence in
  ${\mathrm{UPt}}_{2}{\mathrm{Si}}_{2}$}},\ }\href
  {https://doi.org/10.1103/PhysRevB.102.041112} {\bibfield  {journal} {\bibinfo
   {journal} {Phys. Rev. B}\ }\textbf {\bibinfo {volume} {102}},\ \bibinfo
  {pages} {041112} (\bibinfo {year} {2020})}\BibitemShut {NoStop}%
\bibitem [{\citenamefont {Mydosh}\ \emph {et~al.}(2020)\citenamefont {Mydosh},
  \citenamefont {Oppeneer},\ and\ \citenamefont {Riseborough}}]{Mydosh2020}%
  \BibitemOpen
  \bibfield  {author} {\bibinfo {author} {\bibfnamefont {J.~A.}\ \bibnamefont
  {Mydosh}}, \bibinfo {author} {\bibfnamefont {P.~M.}\ \bibnamefont
  {Oppeneer}},\ and\ \bibinfo {author} {\bibfnamefont {P.~S.}\ \bibnamefont
  {Riseborough}},\ }\bibfield  {title} {\bibinfo {title} {Hidden order and
  beyond: an experimental—theoretical overview of the multifaceted behavior
  of uru2si2},\ }\href {https://doi.org/10.1088/1361-648X/ab5eba} {\bibfield
  {journal} {\bibinfo  {journal} {J. Phys. Condens. Matter}\ }\textbf {\bibinfo
  {volume} {32}},\ \bibinfo {pages} {143002} (\bibinfo {year}
  {2020})}\BibitemShut {NoStop}%
\bibitem [{\citenamefont {Ivanov}\ \emph {et~al.}(2021)\citenamefont {Ivanov},
  \citenamefont {Wan},\ and\ \citenamefont {Savrasov}}]{Ivanov2021}%
  \BibitemOpen
  \bibfield  {author} {\bibinfo {author} {\bibfnamefont {V.}~\bibnamefont
  {Ivanov}}, \bibinfo {author} {\bibfnamefont {X.}~\bibnamefont {Wan}},\ and\
  \bibinfo {author} {\bibfnamefont {S.~Y.}\ \bibnamefont {Savrasov}},\
  }\bibfield  {title} {\bibinfo {title} {{Renormalized quasiparticles,
  topological monopoles, and superconducting line nodes in heavy-fermion
  $\mathrm{Ce}T{X}_{3}$ compounds}},\ }\href
  {https://doi.org/10.1103/PhysRevB.103.L041112} {\bibfield  {journal}
  {\bibinfo  {journal} {Phys. Rev. B}\ }\textbf {\bibinfo {volume} {103}},\
  \bibinfo {pages} {L041112} (\bibinfo {year} {2021})}\BibitemShut {NoStop}%
\bibitem [{\citenamefont {Kundu}\ \emph {et~al.}(2022)\citenamefont {Kundu},
  \citenamefont {Roy}, \citenamefont {Pakhira}, \citenamefont {Wu},
  \citenamefont {Tsujikawa}, \citenamefont {Shirai}, \citenamefont {Johnston},
  \citenamefont {Pasupathy},\ and\ \citenamefont {Valla}}]{Kundu2022}%
  \BibitemOpen
  \bibfield  {author} {\bibinfo {author} {\bibfnamefont {A.~K.}\ \bibnamefont
  {Kundu}}, \bibinfo {author} {\bibfnamefont {T.}~\bibnamefont {Roy}}, \bibinfo
  {author} {\bibfnamefont {S.}~\bibnamefont {Pakhira}}, \bibinfo {author}
  {\bibfnamefont {Z.-B.}\ \bibnamefont {Wu}}, \bibinfo {author} {\bibfnamefont
  {M.}~\bibnamefont {Tsujikawa}}, \bibinfo {author} {\bibfnamefont
  {M.}~\bibnamefont {Shirai}}, \bibinfo {author} {\bibfnamefont {D.~C.}\
  \bibnamefont {Johnston}}, \bibinfo {author} {\bibfnamefont {A.~N.}\
  \bibnamefont {Pasupathy}},\ and\ \bibinfo {author} {\bibfnamefont
  {T.}~\bibnamefont {Valla}},\ }\bibfield  {title} {\bibinfo {title}
  {{Topological electronic structure of YbMg$_2$Bi$_2$ and CaMg$_2$Bi$_2$}},\
  }\href {https://doi.org/10.1038/s41535-022-00474-2} {\bibfield  {journal}
  {\bibinfo  {journal} {npj Quantum Mater.}\ }\textbf {\bibinfo {volume} {7}},\
  \bibinfo {pages} {2397} (\bibinfo {year} {2022})}\BibitemShut {NoStop}%
\bibitem [{\citenamefont {Mallik}\ and\ \citenamefont
  {Sampathkumaran}(1998)}]{Mallik1998}%
  \BibitemOpen
  \bibfield  {author} {\bibinfo {author} {\bibfnamefont {R.}~\bibnamefont
  {Mallik}}\ and\ \bibinfo {author} {\bibfnamefont {E.~V.}\ \bibnamefont
  {Sampathkumaran}},\ }\bibfield  {title} {\bibinfo {title} {{Magnetic
  precursor effects, electrical and magnetoresistance anomalies, and
  heat-capacity behavior of Gd alloys}},\ }\href
  {https://doi.org/10.1103/PhysRevB.58.9178} {\bibfield  {journal} {\bibinfo
  {journal} {Phys. Rev. B}\ }\textbf {\bibinfo {volume} {58}},\ \bibinfo
  {pages} {9178} (\bibinfo {year} {1998})}\BibitemShut {NoStop}%
\bibitem [{\citenamefont {G{\"u}ttler}\ \emph {et~al.}(2016)\citenamefont
  {G{\"u}ttler}, \citenamefont {Generalov}, \citenamefont {Otrokov},
  \citenamefont {Kummer}, \citenamefont {Kliemt}, \citenamefont {Fedorov},
  \citenamefont {Chikina}, \citenamefont {Danzenb{\"a}cher}, \citenamefont
  {Schulz}, \citenamefont {Chulkov}, \citenamefont {Koroteev}, \citenamefont
  {Caroca-Canales}, \citenamefont {Shi}, \citenamefont {Radovic}, \citenamefont
  {Geibel}, \citenamefont {Laubschat}, \citenamefont {Dudin}, \citenamefont
  {Kim}, \citenamefont {Hoesch}, \citenamefont {Krellner},\ and\ \citenamefont
  {Vyalikh}}]{Guttler2016}%
  \BibitemOpen
  \bibfield  {author} {\bibinfo {author} {\bibfnamefont {M.}~\bibnamefont
  {G{\"u}ttler}}, \bibinfo {author} {\bibfnamefont {A.}~\bibnamefont
  {Generalov}}, \bibinfo {author} {\bibfnamefont {M.~M.}\ \bibnamefont
  {Otrokov}}, \bibinfo {author} {\bibfnamefont {K.}~\bibnamefont {Kummer}},
  \bibinfo {author} {\bibfnamefont {K.}~\bibnamefont {Kliemt}}, \bibinfo
  {author} {\bibfnamefont {A.}~\bibnamefont {Fedorov}}, \bibinfo {author}
  {\bibfnamefont {A.}~\bibnamefont {Chikina}}, \bibinfo {author} {\bibfnamefont
  {S.}~\bibnamefont {Danzenb{\"a}cher}}, \bibinfo {author} {\bibfnamefont
  {S.}~\bibnamefont {Schulz}}, \bibinfo {author} {\bibfnamefont {E.~V.}\
  \bibnamefont {Chulkov}}, \bibinfo {author} {\bibfnamefont {Y.~M.}\
  \bibnamefont {Koroteev}}, \bibinfo {author} {\bibfnamefont {N.}~\bibnamefont
  {Caroca-Canales}}, \bibinfo {author} {\bibfnamefont {M.}~\bibnamefont {Shi}},
  \bibinfo {author} {\bibfnamefont {M.}~\bibnamefont {Radovic}}, \bibinfo
  {author} {\bibfnamefont {C.}~\bibnamefont {Geibel}}, \bibinfo {author}
  {\bibfnamefont {C.}~\bibnamefont {Laubschat}}, \bibinfo {author}
  {\bibfnamefont {P.}~\bibnamefont {Dudin}}, \bibinfo {author} {\bibfnamefont
  {T.~K.}\ \bibnamefont {Kim}}, \bibinfo {author} {\bibfnamefont
  {M.}~\bibnamefont {Hoesch}}, \bibinfo {author} {\bibfnamefont
  {C.}~\bibnamefont {Krellner}},\ and\ \bibinfo {author} {\bibfnamefont
  {D.~V.}\ \bibnamefont {Vyalikh}},\ }\bibfield  {title} {\bibinfo {title}
  {{Robust and tunable itinerant ferromagnetism at the silicon surface of the
  antiferromagnet GdRh$_2$Si$_2$}},\ }\href {https://doi.org/10.1038/srep24254}
  {\bibfield  {journal} {\bibinfo  {journal} {Sci. Rep.}\ }\textbf {\bibinfo
  {volume} {6}},\ \bibinfo {pages} {24254} (\bibinfo {year}
  {2016})}\BibitemShut {NoStop}%
\bibitem [{\citenamefont {Garcia}\ \emph {et~al.}(2025)\citenamefont {Garcia},
  \citenamefont {Sereni},\ and\ \citenamefont {Aligia}}]{Garcia2025}%
  \BibitemOpen
  \bibfield  {author} {\bibinfo {author} {\bibfnamefont {D.~J.}\ \bibnamefont
  {Garcia}}, \bibinfo {author} {\bibfnamefont {J.~G.}\ \bibnamefont {Sereni}},\
  and\ \bibinfo {author} {\bibfnamefont {A.~A.}\ \bibnamefont {Aligia}},\
  }\href@noop {} {\bibinfo {title} {{Specific heat of Gd$^{3+}$ and
  Eu$^{2+}$-based magnetic compounds}}} (\bibinfo {year} {2025}),\ \Eprint
  {https://arxiv.org/abs/2410.23519} {arXiv:2410.23519} \BibitemShut {NoStop}%
\bibitem [{\citenamefont {S.i.~Kimura}\ and\ \citenamefont
  {Khim}(2021)}]{Kimura2021}%
  \BibitemOpen
  \bibfield  {author} {\bibinfo {author} {\bibfnamefont {J.~S.}\ \bibnamefont
  {S.i.~Kimura}}\ and\ \bibinfo {author} {\bibfnamefont {S.}~\bibnamefont
  {Khim}},\ }\bibfield  {title} {\bibinfo {title} {{Optical study of the
  electronic structure of locally noncentrosymmetric
  $\mathrm{Ce}{\mathrm{Rh}}_{2}{\mathrm{As}}_{2}$}},\ }\href
  {https://doi.org/10.1103/PhysRevB.104.245116} {\bibfield  {journal} {\bibinfo
   {journal} {Phys. Rev. B}\ }\textbf {\bibinfo {volume} {104}},\ \bibinfo
  {pages} {245116} (\bibinfo {year} {2021})}\BibitemShut {NoStop}%
\bibitem [{\citenamefont {P\"{o}ttgen}\ and\ \citenamefont
  {\L{}\k{a}tka}(2010)}]{Pottgen2010}%
  \BibitemOpen
  \bibfield  {author} {\bibinfo {author} {\bibfnamefont {R.}~\bibnamefont
  {P\"{o}ttgen}}\ and\ \bibinfo {author} {\bibfnamefont {K.}~\bibnamefont
  {\L{}\k{a}tka}},\ }\bibfield  {title} {\bibinfo {title} {{155Gd M{\"o}ssbauer
  Spectroscopy on Intermetallics – An Overview}},\ }\href
  {https://doi.org/https://doi.org/10.1002/zaac.201000171} {\bibfield
  {journal} {\bibinfo  {journal} {Z. Anorg. Allg. Chem.}\ }\textbf {\bibinfo
  {volume} {636}},\ \bibinfo {pages} {2244} (\bibinfo {year}
  {2010})}\BibitemShut {NoStop}%
\bibitem [{\citenamefont {Kliemt}\ \emph {et~al.}(2017)\citenamefont {Kliemt},
  \citenamefont {Hofmann-Kliemt}, \citenamefont {Kummer}, \citenamefont
  {Yakhou-Harris}, \citenamefont {Krellner},\ and\ \citenamefont
  {Geibel}}]{Kliemt2017}%
  \BibitemOpen
  \bibfield  {author} {\bibinfo {author} {\bibfnamefont {K.}~\bibnamefont
  {Kliemt}}, \bibinfo {author} {\bibfnamefont {M.}~\bibnamefont
  {Hofmann-Kliemt}}, \bibinfo {author} {\bibfnamefont {K.}~\bibnamefont
  {Kummer}}, \bibinfo {author} {\bibfnamefont {F.}~\bibnamefont
  {Yakhou-Harris}}, \bibinfo {author} {\bibfnamefont {C.}~\bibnamefont
  {Krellner}},\ and\ \bibinfo {author} {\bibfnamefont {C.}~\bibnamefont
  {Geibel}},\ }\bibfield  {title} {\bibinfo {title}
  {{${\mathrm{GdRh}}_{2}{\mathrm{Si}}_{2}$: An exemplary tetragonal system for
  antiferromagnetic order with weak in-plane anisotropy}},\ }\href
  {https://doi.org/10.1103/PhysRevB.95.134403} {\bibfield  {journal} {\bibinfo
  {journal} {Phys. Rev. B}\ }\textbf {\bibinfo {volume} {95}},\ \bibinfo
  {pages} {134403} (\bibinfo {year} {2017})}\BibitemShut {NoStop}%
\bibitem [{\citenamefont {Barandiaran}\ \emph {et~al.}(1988)\citenamefont
  {Barandiaran}, \citenamefont {Gignoux}, \citenamefont {Schmitt},
  \citenamefont {Gomez-Sal}, \citenamefont {{Rodriguez Fernandez}},
  \citenamefont {Chieux},\ and\ \citenamefont {Schweizer}}]{Barandiaran1988}%
  \BibitemOpen
  \bibfield  {author} {\bibinfo {author} {\bibfnamefont {J.}~\bibnamefont
  {Barandiaran}}, \bibinfo {author} {\bibfnamefont {D.}~\bibnamefont
  {Gignoux}}, \bibinfo {author} {\bibfnamefont {D.}~\bibnamefont {Schmitt}},
  \bibinfo {author} {\bibfnamefont {J.}~\bibnamefont {Gomez-Sal}}, \bibinfo
  {author} {\bibfnamefont {J.}~\bibnamefont {{Rodriguez Fernandez}}}, \bibinfo
  {author} {\bibfnamefont {P.}~\bibnamefont {Chieux}},\ and\ \bibinfo {author}
  {\bibfnamefont {J.}~\bibnamefont {Schweizer}},\ }\bibfield  {title} {\bibinfo
  {title} {{Magnetic properties and magnetic structure of GdNi$_2$Si$_2$ and
  GdCu$_2$Si$_2$ compounds}},\ }\href
  {https://doi.org/https://doi.org/10.1016/0304-8853(88)90299-5} {\bibfield
  {journal} {\bibinfo  {journal} {J. Magn. Magn. Mater.}\ }\textbf {\bibinfo
  {volume} {73}},\ \bibinfo {pages} {233} (\bibinfo {year} {1988})}\BibitemShut
  {NoStop}%
\bibitem [{\citenamefont {Kumar}\ \emph {et~al.}(2007)\citenamefont {Kumar},
  \citenamefont {Singh}, \citenamefont {Suresh}, \citenamefont {Nigam},\ and\
  \citenamefont {Malik}}]{Kumar2007}%
  \BibitemOpen
  \bibfield  {author} {\bibinfo {author} {\bibfnamefont {P.}~\bibnamefont
  {Kumar}}, \bibinfo {author} {\bibfnamefont {N.~K.}\ \bibnamefont {Singh}},
  \bibinfo {author} {\bibfnamefont {K.~G.}\ \bibnamefont {Suresh}}, \bibinfo
  {author} {\bibfnamefont {A.~K.}\ \bibnamefont {Nigam}},\ and\ \bibinfo
  {author} {\bibfnamefont {S.~K.}\ \bibnamefont {Malik}},\ }\bibfield  {title}
  {\bibinfo {title} {{Effect of Ge substitution for Si on the anomalous
  magnetocaloric and magnetoresistance properties of GdMn$_2$Si$_2$
  compounds}},\ }\href {https://doi.org/10.1063/1.2402975} {\bibfield
  {journal} {\bibinfo  {journal} {J. Appl. Phys.}\ }\textbf {\bibinfo {volume}
  {101}},\ \bibinfo {pages} {013908} (\bibinfo {year} {2007})}\BibitemShut
  {NoStop}%
\bibitem [{\citenamefont {Singh}\ \emph {et~al.}(2023)\citenamefont {Singh},
  \citenamefont {Fujishiro}, \citenamefont {Hayami}, \citenamefont {Moody},
  \citenamefont {Nomoto}, \citenamefont {Baral}, \citenamefont {Ukleev},
  \citenamefont {Cubitt}, \citenamefont {Steinke}, \citenamefont {Gawryluk},
  \citenamefont {Pomjakushina}, \citenamefont {{\={O}}nuki}, \citenamefont
  {Arita}, \citenamefont {Tokura}, \citenamefont {Kanazawa},\ and\
  \citenamefont {White}}]{Singh2023}%
  \BibitemOpen
  \bibfield  {author} {\bibinfo {author} {\bibfnamefont {D.}~\bibnamefont
  {Singh}}, \bibinfo {author} {\bibfnamefont {Y.}~\bibnamefont {Fujishiro}},
  \bibinfo {author} {\bibfnamefont {S.}~\bibnamefont {Hayami}}, \bibinfo
  {author} {\bibfnamefont {S.~H.}\ \bibnamefont {Moody}}, \bibinfo {author}
  {\bibfnamefont {T.}~\bibnamefont {Nomoto}}, \bibinfo {author} {\bibfnamefont
  {P.~R.}\ \bibnamefont {Baral}}, \bibinfo {author} {\bibfnamefont
  {V.}~\bibnamefont {Ukleev}}, \bibinfo {author} {\bibfnamefont
  {R.}~\bibnamefont {Cubitt}}, \bibinfo {author} {\bibfnamefont {N.-J.}\
  \bibnamefont {Steinke}}, \bibinfo {author} {\bibfnamefont {D.~J.}\
  \bibnamefont {Gawryluk}}, \bibinfo {author} {\bibfnamefont {E.}~\bibnamefont
  {Pomjakushina}}, \bibinfo {author} {\bibfnamefont {Y.}~\bibnamefont
  {{\={O}}nuki}}, \bibinfo {author} {\bibfnamefont {R.}~\bibnamefont {Arita}},
  \bibinfo {author} {\bibfnamefont {Y.}~\bibnamefont {Tokura}}, \bibinfo
  {author} {\bibfnamefont {N.}~\bibnamefont {Kanazawa}},\ and\ \bibinfo
  {author} {\bibfnamefont {J.~S.}\ \bibnamefont {White}},\ }\bibfield  {title}
  {\bibinfo {title} {{Transition between distinct hybrid skyrmion textures
  through their hexagonal-to-square crystal transformation in a polar
  magnet}},\ }\href {https://doi.org/10.1038/s41467-023-43814-x} {\bibfield
  {journal} {\bibinfo  {journal} {Nat. Commun.}\ }\textbf {\bibinfo {volume}
  {14}},\ \bibinfo {pages} {8050} (\bibinfo {year} {2023})}\BibitemShut
  {NoStop}%
\bibitem [{\citenamefont {Matsumura}\ \emph {et~al.}(2024)\citenamefont
  {Matsumura}, \citenamefont {Kurauchi}, \citenamefont {Tsukagoshi},
  \citenamefont {Higa}, \citenamefont {Nakao}, \citenamefont {Kakihana},
  \citenamefont {Hedo}, \citenamefont {Nakama},\ and\ \citenamefont
  {Ōnuki}}]{Matsumura2024}%
  \BibitemOpen
  \bibfield  {author} {\bibinfo {author} {\bibfnamefont {T.}~\bibnamefont
  {Matsumura}}, \bibinfo {author} {\bibfnamefont {K.}~\bibnamefont {Kurauchi}},
  \bibinfo {author} {\bibfnamefont {M.}~\bibnamefont {Tsukagoshi}}, \bibinfo
  {author} {\bibfnamefont {N.}~\bibnamefont {Higa}}, \bibinfo {author}
  {\bibfnamefont {H.}~\bibnamefont {Nakao}}, \bibinfo {author} {\bibfnamefont
  {M.}~\bibnamefont {Kakihana}}, \bibinfo {author} {\bibfnamefont
  {M.}~\bibnamefont {Hedo}}, \bibinfo {author} {\bibfnamefont {T.}~\bibnamefont
  {Nakama}},\ and\ \bibinfo {author} {\bibfnamefont {Y.}~\bibnamefont
  {Ōnuki}},\ }\bibfield  {title} {\bibinfo {title} {{Helicity Unification by
  Triangular Skyrmion Lattice Formation in the Noncentrosymmetric Tetragonal
  Magnet EuNiGe$_3$}},\ }\href {https://doi.org/10.7566/JPSJ.93.074705}
  {\bibfield  {journal} {\bibinfo  {journal} {J. Phys. Soc. Jpn.}\ }\textbf
  {\bibinfo {volume} {93}},\ \bibinfo {pages} {074705} (\bibinfo {year}
  {2024})}\BibitemShut {NoStop}%
\bibitem [{\citenamefont {Yasui}\ \emph {et~al.}(2020)\citenamefont {Yasui},
  \citenamefont {Butler}, \citenamefont {Khanh}, \citenamefont {Hayami},
  \citenamefont {Nomoto}, \citenamefont {Hanaguri}, \citenamefont {Motome},
  \citenamefont {Arita}, \citenamefont {Arima}, \citenamefont {Tokura},\ and\
  \citenamefont {Seki}}]{Yasui2020}%
  \BibitemOpen
  \bibfield  {author} {\bibinfo {author} {\bibfnamefont {Y.}~\bibnamefont
  {Yasui}}, \bibinfo {author} {\bibfnamefont {C.~J.}\ \bibnamefont {Butler}},
  \bibinfo {author} {\bibfnamefont {N.~D.}\ \bibnamefont {Khanh}}, \bibinfo
  {author} {\bibfnamefont {S.}~\bibnamefont {Hayami}}, \bibinfo {author}
  {\bibfnamefont {T.}~\bibnamefont {Nomoto}}, \bibinfo {author} {\bibfnamefont
  {T.}~\bibnamefont {Hanaguri}}, \bibinfo {author} {\bibfnamefont
  {Y.}~\bibnamefont {Motome}}, \bibinfo {author} {\bibfnamefont
  {R.}~\bibnamefont {Arita}}, \bibinfo {author} {\bibfnamefont {T.-h.}\
  \bibnamefont {Arima}}, \bibinfo {author} {\bibfnamefont {Y.}~\bibnamefont
  {Tokura}},\ and\ \bibinfo {author} {\bibfnamefont {S.}~\bibnamefont {Seki}},\
  }\bibfield  {title} {\bibinfo {title} {{Imaging the coupling between
  itinerant electrons and localised moments in the centrosymmetric skyrmion
  magnet GdRu$_2$Si$_2$}},\ }\href {https://doi.org/10.1038/s41467-020-19751-4}
  {\bibfield  {journal} {\bibinfo  {journal} {Nat. Commun.}\ }\textbf {\bibinfo
  {volume} {11}},\ \bibinfo {pages} {5925} (\bibinfo {year}
  {2020})}\BibitemShut {NoStop}%
\bibitem [{\citenamefont {Wood}\ \emph {et~al.}(2023)\citenamefont {Wood},
  \citenamefont {Khalyavin}, \citenamefont {Mayoh}, \citenamefont {Bouaziz},
  \citenamefont {Hall}, \citenamefont {Holt}, \citenamefont {Orlandi},
  \citenamefont {Manuel}, \citenamefont {Bl\"ugel}, \citenamefont {Staunton},
  \citenamefont {Petrenko}, \citenamefont {Lees},\ and\ \citenamefont
  {Balakrishnan}}]{Wood2023}%
  \BibitemOpen
  \bibfield  {author} {\bibinfo {author} {\bibfnamefont {G.~D.~A.}\
  \bibnamefont {Wood}}, \bibinfo {author} {\bibfnamefont {D.~D.}\ \bibnamefont
  {Khalyavin}}, \bibinfo {author} {\bibfnamefont {D.~A.}\ \bibnamefont
  {Mayoh}}, \bibinfo {author} {\bibfnamefont {J.}~\bibnamefont {Bouaziz}},
  \bibinfo {author} {\bibfnamefont {A.~E.}\ \bibnamefont {Hall}}, \bibinfo
  {author} {\bibfnamefont {S.~J.~R.}\ \bibnamefont {Holt}}, \bibinfo {author}
  {\bibfnamefont {F.}~\bibnamefont {Orlandi}}, \bibinfo {author} {\bibfnamefont
  {P.}~\bibnamefont {Manuel}}, \bibinfo {author} {\bibfnamefont
  {S.}~\bibnamefont {Bl\"ugel}}, \bibinfo {author} {\bibfnamefont {J.~B.}\
  \bibnamefont {Staunton}}, \bibinfo {author} {\bibfnamefont {O.~A.}\
  \bibnamefont {Petrenko}}, \bibinfo {author} {\bibfnamefont {M.~R.}\
  \bibnamefont {Lees}},\ and\ \bibinfo {author} {\bibfnamefont
  {G.}~\bibnamefont {Balakrishnan}},\ }\bibfield  {title} {\bibinfo {title}
  {{Double-$Q$ ground state with topological charge stripes in the
  centrosymmetric skyrmion candidate ${\mathrm{GdRu}}_{2}{\mathrm{Si}}_{2}$}},\
  }\href {https://doi.org/10.1103/PhysRevB.107.L180402} {\bibfield  {journal}
  {\bibinfo  {journal} {Phys. Rev. B}\ }\textbf {\bibinfo {volume} {107}},\
  \bibinfo {pages} {L180402} (\bibinfo {year} {2023})}\BibitemShut {NoStop}%
\bibitem [{\citenamefont {Bouaziz}\ \emph {et~al.}(2022)\citenamefont
  {Bouaziz}, \citenamefont {Mendive-Tapia}, \citenamefont {Bl\"ugel},\ and\
  \citenamefont {Staunton}}]{Bouaziz2022}%
  \BibitemOpen
  \bibfield  {author} {\bibinfo {author} {\bibfnamefont {J.}~\bibnamefont
  {Bouaziz}}, \bibinfo {author} {\bibfnamefont {E.}~\bibnamefont
  {Mendive-Tapia}}, \bibinfo {author} {\bibfnamefont {S.}~\bibnamefont
  {Bl\"ugel}},\ and\ \bibinfo {author} {\bibfnamefont {J.~B.}\ \bibnamefont
  {Staunton}},\ }\bibfield  {title} {\bibinfo {title} {{Fermi-Surface Origin of
  Skyrmion Lattices in Centrosymmetric Rare-Earth Intermetallics}},\ }\href
  {https://doi.org/10.1103/PhysRevLett.128.157206} {\bibfield  {journal}
  {\bibinfo  {journal} {Phys. Rev. Lett.}\ }\textbf {\bibinfo {volume} {128}},\
  \bibinfo {pages} {157206} (\bibinfo {year} {2022})}\BibitemShut {NoStop}%
\bibitem [{\citenamefont {Kurumaji}\ \emph {et~al.}(2019)\citenamefont
  {Kurumaji}, \citenamefont {Nakajima}, \citenamefont {Hirschberger},
  \citenamefont {Kikkawa}, \citenamefont {Yamasaki}, \citenamefont {Sagayama},
  \citenamefont {Nakao}, \citenamefont {Taguchi}, \citenamefont {Arima},\ and\
  \citenamefont {Tokura}}]{Kurumaji2019}%
  \BibitemOpen
  \bibfield  {author} {\bibinfo {author} {\bibfnamefont {T.}~\bibnamefont
  {Kurumaji}}, \bibinfo {author} {\bibfnamefont {T.}~\bibnamefont {Nakajima}},
  \bibinfo {author} {\bibfnamefont {M.}~\bibnamefont {Hirschberger}}, \bibinfo
  {author} {\bibfnamefont {A.}~\bibnamefont {Kikkawa}}, \bibinfo {author}
  {\bibfnamefont {Y.}~\bibnamefont {Yamasaki}}, \bibinfo {author}
  {\bibfnamefont {H.}~\bibnamefont {Sagayama}}, \bibinfo {author}
  {\bibfnamefont {H.}~\bibnamefont {Nakao}}, \bibinfo {author} {\bibfnamefont
  {Y.}~\bibnamefont {Taguchi}}, \bibinfo {author} {\bibfnamefont {T.-H.}\
  \bibnamefont {Arima}},\ and\ \bibinfo {author} {\bibfnamefont
  {Y.}~\bibnamefont {Tokura}},\ }\bibfield  {title} {\bibinfo {title}
  {{Skyrmion lattice with a giant topological Hall effect in a frustrated
  triangular-lattice magnet}},\ }\href
  {https://doi.org/10.1126/science.aau0968} {\bibfield  {journal} {\bibinfo
  {journal} {Science}\ }\textbf {\bibinfo {volume} {365}},\ \bibinfo {pages}
  {914} (\bibinfo {year} {2019})}\BibitemShut {NoStop}%
\bibitem [{\citenamefont {Momma}\ and\ \citenamefont
  {Izumi}(2008)}]{Momma2008}%
  \BibitemOpen
  \bibfield  {author} {\bibinfo {author} {\bibfnamefont {K.}~\bibnamefont
  {Momma}}\ and\ \bibinfo {author} {\bibfnamefont {F.}~\bibnamefont {Izumi}},\
  }\bibfield  {title} {\bibinfo {title} {{{\it VESTA}: a three-dimensional
  visualization system for electronic and structural analysis}},\ }\href
  {https://doi.org/10.1107/S0021889808012016} {\bibfield  {journal} {\bibinfo
  {journal} {J. Appl. Crystallogr.}\ }\textbf {\bibinfo {volume} {41}},\
  \bibinfo {pages} {653} (\bibinfo {year} {2008})}\BibitemShut {NoStop}%
\bibitem [{SM()}]{SM}%
  \BibitemOpen
  \href@noop {} {\bibinfo {title} {See supplemental material at [url will be
  inserted by publisher] for the cif files, decomposed optical spectra at other
  temperatures, and additional computational results.}}\BibitemShut {Stop}%
\bibitem [{\citenamefont {Mulder}\ \emph {et~al.}(1993)\citenamefont {Mulder},
  \citenamefont {Thiel},\ and\ \citenamefont {Buschow}}]{Mulder1993}%
  \BibitemOpen
  \bibfield  {author} {\bibinfo {author} {\bibfnamefont {F.}~\bibnamefont
  {Mulder}}, \bibinfo {author} {\bibfnamefont {R.}~\bibnamefont {Thiel}},\ and\
  \bibinfo {author} {\bibfnamefont {K.}~\bibnamefont {Buschow}},\ }\bibfield
  {title} {\bibinfo {title} {{$^{155}$Gd M\"ossbauer effect and magnetic
  properties of ternary rare earth compounds of the type RT$_2$Ge$_2$ (T=3d,
  4d)}},\ }\href {https://doi.org/https://doi.org/10.1016/0925-8388(93)90512-L}
  {\bibfield  {journal} {\bibinfo  {journal} {J. Alloys Compd.}\ }\textbf
  {\bibinfo {volume} {202}},\ \bibinfo {pages} {29} (\bibinfo {year}
  {1993})}\BibitemShut {NoStop}%
\bibitem [{\citenamefont {Mulder}\ \emph {et~al.}(1994)\citenamefont {Mulder},
  \citenamefont {Thiel},\ and\ \citenamefont {Buschow}}]{Mulder1994}%
  \BibitemOpen
  \bibfield  {author} {\bibinfo {author} {\bibfnamefont {F.}~\bibnamefont
  {Mulder}}, \bibinfo {author} {\bibfnamefont {R.}~\bibnamefont {Thiel}},\ and\
  \bibinfo {author} {\bibfnamefont {K.}~\bibnamefont {Buschow}},\ }\bibfield
  {title} {\bibinfo {title} {{$^{155}$Gd M\"ossbauer effect in several
  BaNiSn$_3$-type compounds}},\ }\href
  {https://doi.org/https://doi.org/10.1016/0925-8388(94)91048-0} {\bibfield
  {journal} {\bibinfo  {journal} {J. Alloys Compd.}\ }\textbf {\bibinfo
  {volume} {216}},\ \bibinfo {pages} {95} (\bibinfo {year} {1994})}\BibitemShut
  {NoStop}%
\bibitem [{\citenamefont {Homes}\ \emph {et~al.}(1993)\citenamefont {Homes},
  \citenamefont {Reedyk}, \citenamefont {Cradles},\ and\ \citenamefont
  {Timusk}}]{Homes1993}%
  \BibitemOpen
  \bibfield  {author} {\bibinfo {author} {\bibfnamefont {C.~C.}\ \bibnamefont
  {Homes}}, \bibinfo {author} {\bibfnamefont {M.}~\bibnamefont {Reedyk}},
  \bibinfo {author} {\bibfnamefont {D.~A.}\ \bibnamefont {Cradles}},\ and\
  \bibinfo {author} {\bibfnamefont {T.}~\bibnamefont {Timusk}},\ }\bibfield
  {title} {\bibinfo {title} {{Technique for measuring the reflectance of
  irregular, submillimeter-sized samples}},\ }\href
  {https://doi.org/10.1364/AO.32.002976} {\bibfield  {journal} {\bibinfo
  {journal} {Appl. Opt.}\ }\textbf {\bibinfo {volume} {32}},\ \bibinfo {pages}
  {2976} (\bibinfo {year} {1993})}\BibitemShut {NoStop}%
\bibitem [{\citenamefont {Tanner}(2015)}]{Tanner2015}%
  \BibitemOpen
  \bibfield  {author} {\bibinfo {author} {\bibfnamefont {D.~B.}\ \bibnamefont
  {Tanner}},\ }\bibfield  {title} {\bibinfo {title} {{Use of x-ray scattering
  functions in Kramers-Kronig analysis of reflectance}},\ }\href
  {https://doi.org/10.1103/PhysRevB.91.035123} {\bibfield  {journal} {\bibinfo
  {journal} {Phys. Rev. B}\ }\textbf {\bibinfo {volume} {91}},\ \bibinfo
  {pages} {035123} (\bibinfo {year} {2015})}\BibitemShut {NoStop}%
\bibitem [{\citenamefont {Dressel}\ and\ \citenamefont
  {Gr\"uner}(2002)}]{Dressel2002}%
  \BibitemOpen
  \bibfield  {author} {\bibinfo {author} {\bibfnamefont {M.}~\bibnamefont
  {Dressel}}\ and\ \bibinfo {author} {\bibfnamefont {G.}~\bibnamefont
  {Gr\"uner}},\ }\href {https://doi.org/10.1017/CBO9780511606168} {\emph
  {\bibinfo {title} {{Electrodynamics of Solids: Optical Properties of
  Electrons in Matter}}}}\ (\bibinfo  {publisher} {Cambridge University
  Press},\ \bibinfo {year} {2002})\BibitemShut {NoStop}%
\bibitem [{\citenamefont {Blaha}\ \emph {et~al.}()\citenamefont {Blaha},
  \citenamefont {Schwarz}, \citenamefont {Madsen}, \citenamefont {Kvasnicka},
  \citenamefont {Luitz}, \citenamefont {Laskowski}, \citenamefont {Tran},\ and\
  \citenamefont {Marks}}]{wien2k}%
  \BibitemOpen
  \bibfield  {author} {\bibinfo {author} {\bibfnamefont {P.}~\bibnamefont
  {Blaha}}, \bibinfo {author} {\bibfnamefont {K.}~\bibnamefont {Schwarz}},
  \bibinfo {author} {\bibfnamefont {G.}~\bibnamefont {Madsen}}, \bibinfo
  {author} {\bibfnamefont {D.}~\bibnamefont {Kvasnicka}}, \bibinfo {author}
  {\bibfnamefont {J.}~\bibnamefont {Luitz}}, \bibinfo {author} {\bibfnamefont
  {R.}~\bibnamefont {Laskowski}}, \bibinfo {author} {\bibfnamefont
  {F.}~\bibnamefont {Tran}},\ and\ \bibinfo {author} {\bibfnamefont
  {L.}~\bibnamefont {Marks}},\ }\href@noop {} {}\bibinfo {note} {WIEN2k, An
  Augmented Plane Wave + Local Orbitals Program for Calculating Crystal
  Properties (Karlheinz Schwarz, Techn. Universit\"at Wien, Austria), 2018.
  ISBN 3-9501031-1-2}\BibitemShut {NoStop}%
\bibitem [{\citenamefont {Blaha}\ \emph {et~al.}(2020)\citenamefont {Blaha},
  \citenamefont {Schwarz}, \citenamefont {Tran}, \citenamefont {Laskowski},
  \citenamefont {Madsen},\ and\ \citenamefont {Marks}}]{Blaha2020}%
  \BibitemOpen
  \bibfield  {author} {\bibinfo {author} {\bibfnamefont {P.}~\bibnamefont
  {Blaha}}, \bibinfo {author} {\bibfnamefont {K.}~\bibnamefont {Schwarz}},
  \bibinfo {author} {\bibfnamefont {F.}~\bibnamefont {Tran}}, \bibinfo {author}
  {\bibfnamefont {R.}~\bibnamefont {Laskowski}}, \bibinfo {author}
  {\bibfnamefont {G.~K.~H.}\ \bibnamefont {Madsen}},\ and\ \bibinfo {author}
  {\bibfnamefont {L.~D.}\ \bibnamefont {Marks}},\ }\bibfield  {title} {\bibinfo
  {title} {{WIEN2k: An APW+lo program for calculating the properties of
  solids}},\ }\href {https://doi.org/10.1063/1.5143061} {\bibfield  {journal}
  {\bibinfo  {journal} {J. Chem. Phys.}\ }\textbf {\bibinfo {volume} {152}},\
  \bibinfo {pages} {074101} (\bibinfo {year} {2020})}\BibitemShut {NoStop}%
\bibitem [{\citenamefont {Perdew}\ \emph {et~al.}(1996)\citenamefont {Perdew},
  \citenamefont {Burke},\ and\ \citenamefont {Ernzerhof}}]{pbe96}%
  \BibitemOpen
  \bibfield  {author} {\bibinfo {author} {\bibfnamefont {J.~P.}\ \bibnamefont
  {Perdew}}, \bibinfo {author} {\bibfnamefont {K.}~\bibnamefont {Burke}},\ and\
  \bibinfo {author} {\bibfnamefont {M.}~\bibnamefont {Ernzerhof}},\ }\bibfield
  {title} {\bibinfo {title} {{Generalized Gradient Approximation Made
  Simple}},\ }\href {https://doi.org/10.1103/PhysRevLett.77.3865} {\bibfield
  {journal} {\bibinfo  {journal} {Phys. Rev. Lett.}\ }\textbf {\bibinfo
  {volume} {77}},\ \bibinfo {pages} {3865} (\bibinfo {year}
  {1996})}\BibitemShut {NoStop}%
\bibitem [{\citenamefont {Ambrosch-Draxl}\ and\ \citenamefont
  {Sofo}(2006)}]{Draxl2006}%
  \BibitemOpen
  \bibfield  {author} {\bibinfo {author} {\bibfnamefont {C.}~\bibnamefont
  {Ambrosch-Draxl}}\ and\ \bibinfo {author} {\bibfnamefont {J.~O.}\
  \bibnamefont {Sofo}},\ }\bibfield  {title} {\bibinfo {title} {{Linear optical
  properties of solids within the full-potential linearized augmented planewave
  method}},\ }\href {https://doi.org/https://doi.org/10.1016/j.cpc.2006.03.005}
  {\bibfield  {journal} {\bibinfo  {journal} {Comput. Phys. Commun.}\ }\textbf
  {\bibinfo {volume} {175}},\ \bibinfo {pages} {1} (\bibinfo {year}
  {2006})}\BibitemShut {NoStop}%
\bibitem [{\citenamefont {Schilling}\ \emph {et~al.}(2017)\citenamefont
  {Schilling}, \citenamefont {L\"ohle}, \citenamefont {Neubauer}, \citenamefont
  {Shekhar}, \citenamefont {Felser}, \citenamefont {Dressel},\ and\
  \citenamefont {Pronin}}]{Schilling2017}%
  \BibitemOpen
  \bibfield  {author} {\bibinfo {author} {\bibfnamefont {M.~B.}\ \bibnamefont
  {Schilling}}, \bibinfo {author} {\bibfnamefont {A.}~\bibnamefont {L\"ohle}},
  \bibinfo {author} {\bibfnamefont {D.}~\bibnamefont {Neubauer}}, \bibinfo
  {author} {\bibfnamefont {C.}~\bibnamefont {Shekhar}}, \bibinfo {author}
  {\bibfnamefont {C.}~\bibnamefont {Felser}}, \bibinfo {author} {\bibfnamefont
  {M.}~\bibnamefont {Dressel}},\ and\ \bibinfo {author} {\bibfnamefont {A.~V.}\
  \bibnamefont {Pronin}},\ }\bibfield  {title} {\bibinfo {title} {{Two-channel
  conduction in YbPtBi}},\ }\href {https://doi.org/10.1103/PhysRevB.95.155201}
  {\bibfield  {journal} {\bibinfo  {journal} {Phys. Rev. B}\ }\textbf {\bibinfo
  {volume} {95}},\ \bibinfo {pages} {155201} (\bibinfo {year}
  {2017})}\BibitemShut {NoStop}%
\bibitem [{\citenamefont {Neubauer}\ \emph {et~al.}(2018)\citenamefont
  {Neubauer}, \citenamefont {Yaresko}, \citenamefont {Li}, \citenamefont
  {L\"ohle}, \citenamefont {H\"ubner}, \citenamefont {Schilling}, \citenamefont
  {Shekhar}, \citenamefont {Felser}, \citenamefont {Dressel},\ and\
  \citenamefont {Pronin}}]{Neubauer2018}%
  \BibitemOpen
  \bibfield  {author} {\bibinfo {author} {\bibfnamefont {D.}~\bibnamefont
  {Neubauer}}, \bibinfo {author} {\bibfnamefont {A.}~\bibnamefont {Yaresko}},
  \bibinfo {author} {\bibfnamefont {W.}~\bibnamefont {Li}}, \bibinfo {author}
  {\bibfnamefont {A.}~\bibnamefont {L\"ohle}}, \bibinfo {author} {\bibfnamefont
  {R.}~\bibnamefont {H\"ubner}}, \bibinfo {author} {\bibfnamefont {M.~B.}\
  \bibnamefont {Schilling}}, \bibinfo {author} {\bibfnamefont {C.}~\bibnamefont
  {Shekhar}}, \bibinfo {author} {\bibfnamefont {C.}~\bibnamefont {Felser}},
  \bibinfo {author} {\bibfnamefont {M.}~\bibnamefont {Dressel}},\ and\ \bibinfo
  {author} {\bibfnamefont {A.~V.}\ \bibnamefont {Pronin}},\ }\bibfield  {title}
  {\bibinfo {title} {{Optical conductivity of the Weyl semimetal NbP}},\ }\href
  {https://doi.org/10.1103/PhysRevB.98.195203} {\bibfield  {journal} {\bibinfo
  {journal} {Phys. Rev. B}\ }\textbf {\bibinfo {volume} {98}},\ \bibinfo
  {pages} {195203} (\bibinfo {year} {2018})}\BibitemShut {NoStop}%
\bibitem [{\citenamefont {Yang}\ \emph {et~al.}(2024)\citenamefont {Yang},
  \citenamefont {Le}, \citenamefont {Zhu}, \citenamefont {Wang}, \citenamefont
  {Shang}, \citenamefont {Dai}, \citenamefont {Hu},\ and\ \citenamefont
  {Dressel}}]{Yang2024}%
  \BibitemOpen
  \bibfield  {author} {\bibinfo {author} {\bibfnamefont {R.}~\bibnamefont
  {Yang}}, \bibinfo {author} {\bibfnamefont {C.-C.}\ \bibnamefont {Le}},
  \bibinfo {author} {\bibfnamefont {P.}~\bibnamefont {Zhu}}, \bibinfo {author}
  {\bibfnamefont {Z.-W.}\ \bibnamefont {Wang}}, \bibinfo {author}
  {\bibfnamefont {T.}~\bibnamefont {Shang}}, \bibinfo {author} {\bibfnamefont
  {Y.-M.}\ \bibnamefont {Dai}}, \bibinfo {author} {\bibfnamefont {J.-P.}\
  \bibnamefont {Hu}},\ and\ \bibinfo {author} {\bibfnamefont {M.}~\bibnamefont
  {Dressel}},\ }\bibfield  {title} {\bibinfo {title} {{Charge density wave
  transition in the magnetic topological semimetal ${\mathrm{EuAl}}_{4}$}},\
  }\href {https://doi.org/10.1103/PhysRevB.109.L041113} {\bibfield  {journal}
  {\bibinfo  {journal} {Phys. Rev. B}\ }\textbf {\bibinfo {volume} {109}},\
  \bibinfo {pages} {L041113} (\bibinfo {year} {2024})}\BibitemShut {NoStop}%
\bibitem [{\citenamefont {Homes}\ \emph {et~al.}(2020)\citenamefont {Homes},
  \citenamefont {Wolf},\ and\ \citenamefont {Meingast}}]{Homes2020}%
  \BibitemOpen
  \bibfield  {author} {\bibinfo {author} {\bibfnamefont {C.~C.}\ \bibnamefont
  {Homes}}, \bibinfo {author} {\bibfnamefont {T.}~\bibnamefont {Wolf}},\ and\
  \bibinfo {author} {\bibfnamefont {C.}~\bibnamefont {Meingast}},\ }\bibfield
  {title} {\bibinfo {title} {{Anisotropic optical properties of detwinned
  ${\mathrm{BaFe}}_{2}{\mathrm{As}}_{2}$}},\ }\href
  {https://doi.org/10.1103/PhysRevB.102.155135} {\bibfield  {journal} {\bibinfo
   {journal} {Phys. Rev. B}\ }\textbf {\bibinfo {volume} {102}},\ \bibinfo
  {pages} {155135} (\bibinfo {year} {2020})}\BibitemShut {NoStop}%
\bibitem [{\citenamefont {Nakajima}\ \emph {et~al.}(2014)\citenamefont
  {Nakajima}, \citenamefont {Ishida}, \citenamefont {Tanaka}, \citenamefont
  {Kihou}, \citenamefont {Tomioka}, \citenamefont {Saito}, \citenamefont {Lee},
  \citenamefont {Fukazawa}, \citenamefont {Kohori}, \citenamefont {Kakeshita},
  \citenamefont {Iyo}, \citenamefont {Ito}, \citenamefont {Eisaki},\ and\
  \citenamefont {Uchida}}]{Nakajima2014}%
  \BibitemOpen
  \bibfield  {author} {\bibinfo {author} {\bibfnamefont {M.}~\bibnamefont
  {Nakajima}}, \bibinfo {author} {\bibfnamefont {S.}~\bibnamefont {Ishida}},
  \bibinfo {author} {\bibfnamefont {T.}~\bibnamefont {Tanaka}}, \bibinfo
  {author} {\bibfnamefont {K.}~\bibnamefont {Kihou}}, \bibinfo {author}
  {\bibfnamefont {Y.}~\bibnamefont {Tomioka}}, \bibinfo {author} {\bibfnamefont
  {T.}~\bibnamefont {Saito}}, \bibinfo {author} {\bibfnamefont {C.-H.}\
  \bibnamefont {Lee}}, \bibinfo {author} {\bibfnamefont {H.}~\bibnamefont
  {Fukazawa}}, \bibinfo {author} {\bibfnamefont {Y.}~\bibnamefont {Kohori}},
  \bibinfo {author} {\bibfnamefont {T.}~\bibnamefont {Kakeshita}}, \bibinfo
  {author} {\bibfnamefont {A.}~\bibnamefont {Iyo}}, \bibinfo {author}
  {\bibfnamefont {T.}~\bibnamefont {Ito}}, \bibinfo {author} {\bibfnamefont
  {H.}~\bibnamefont {Eisaki}},\ and\ \bibinfo {author} {\bibfnamefont {S.-I.}\
  \bibnamefont {Uchida}},\ }\bibfield  {title} {\bibinfo {title} {{Strong
  Electronic Correlations in Iron Pnictides: Comparison of Optical Spectra for
  BaFe$_2$As$_2$-Related Compounds}},\ }\href
  {https://doi.org/10.7566/JPSJ.83.104703} {\bibfield  {journal} {\bibinfo
  {journal} {J. Phys. Soc. Jpn.}\ }\textbf {\bibinfo {volume} {83}},\ \bibinfo
  {pages} {104703} (\bibinfo {year} {2014})}\BibitemShut {NoStop}%
\bibitem [{\citenamefont {Maulana}\ \emph {et~al.}(2020)\citenamefont
  {Maulana}, \citenamefont {Manna}, \citenamefont {Uykur}, \citenamefont
  {Felser}, \citenamefont {Dressel},\ and\ \citenamefont
  {Pronin}}]{Maulana2020}%
  \BibitemOpen
  \bibfield  {author} {\bibinfo {author} {\bibfnamefont {L.~Z.}\ \bibnamefont
  {Maulana}}, \bibinfo {author} {\bibfnamefont {K.}~\bibnamefont {Manna}},
  \bibinfo {author} {\bibfnamefont {E.}~\bibnamefont {Uykur}}, \bibinfo
  {author} {\bibfnamefont {C.}~\bibnamefont {Felser}}, \bibinfo {author}
  {\bibfnamefont {M.}~\bibnamefont {Dressel}},\ and\ \bibinfo {author}
  {\bibfnamefont {A.~V.}\ \bibnamefont {Pronin}},\ }\bibfield  {title}
  {\bibinfo {title} {{Optical conductivity of multifold fermions: The case of
  RhSi}},\ }\href {https://doi.org/10.1103/PhysRevResearch.2.023018} {\bibfield
   {journal} {\bibinfo  {journal} {Phys. Rev. Res.}\ }\textbf {\bibinfo
  {volume} {2}},\ \bibinfo {pages} {023018} (\bibinfo {year}
  {2020})}\BibitemShut {NoStop}%
\bibitem [{\citenamefont {Bari\ifmmode \check{s}\else
  \v{s}\fi{}i\ifmmode~\acute{c}\else \'{c}\fi{}}\ \emph
  {et~al.}(2010)\citenamefont {Bari\ifmmode \check{s}\else
  \v{s}\fi{}i\ifmmode~\acute{c}\else \'{c}\fi{}}, \citenamefont {Wu},
  \citenamefont {Dressel}, \citenamefont {Li}, \citenamefont {Cao},\ and\
  \citenamefont {Xu}}]{Barisic2010}%
  \BibitemOpen
  \bibfield  {author} {\bibinfo {author} {\bibfnamefont {N.}~\bibnamefont
  {Bari\ifmmode \check{s}\else \v{s}\fi{}i\ifmmode~\acute{c}\else \'{c}\fi{}}},
  \bibinfo {author} {\bibfnamefont {D.}~\bibnamefont {Wu}}, \bibinfo {author}
  {\bibfnamefont {M.}~\bibnamefont {Dressel}}, \bibinfo {author} {\bibfnamefont
  {L.~J.}\ \bibnamefont {Li}}, \bibinfo {author} {\bibfnamefont {G.~H.}\
  \bibnamefont {Cao}},\ and\ \bibinfo {author} {\bibfnamefont {Z.~A.}\
  \bibnamefont {Xu}},\ }\bibfield  {title} {\bibinfo {title} {{Electrodynamics
  of electron-doped iron pnictide superconductors: Normal-state properties}},\
  }\href {https://doi.org/10.1103/PhysRevB.82.054518} {\bibfield  {journal}
  {\bibinfo  {journal} {Phys. Rev. B}\ }\textbf {\bibinfo {volume} {82}},\
  \bibinfo {pages} {054518} (\bibinfo {year} {2010})}\BibitemShut {NoStop}%
\bibitem [{\citenamefont {Kemmler}\ \emph {et~al.}(2018)\citenamefont
  {Kemmler}, \citenamefont {Hübner}, \citenamefont {Löhle}, \citenamefont
  {Neubauer}, \citenamefont {Voloshenko}, \citenamefont {Schoop}, \citenamefont
  {Dressel},\ and\ \citenamefont {Pronin}}]{Kemmler2018}%
  \BibitemOpen
  \bibfield  {author} {\bibinfo {author} {\bibfnamefont {R.}~\bibnamefont
  {Kemmler}}, \bibinfo {author} {\bibfnamefont {R.}~\bibnamefont {Hübner}},
  \bibinfo {author} {\bibfnamefont {A.}~\bibnamefont {Löhle}}, \bibinfo
  {author} {\bibfnamefont {D.}~\bibnamefont {Neubauer}}, \bibinfo {author}
  {\bibfnamefont {I.}~\bibnamefont {Voloshenko}}, \bibinfo {author}
  {\bibfnamefont {L.~M.}\ \bibnamefont {Schoop}}, \bibinfo {author}
  {\bibfnamefont {M.}~\bibnamefont {Dressel}},\ and\ \bibinfo {author}
  {\bibfnamefont {A.~V.}\ \bibnamefont {Pronin}},\ }\bibfield  {title}
  {\bibinfo {title} {{Free-carrier dynamics in Au$_2$Pb probed by optical
  conductivity measurements}},\ }\href
  {https://doi.org/10.1088/1361-648X/aae7b1} {\bibfield  {journal} {\bibinfo
  {journal} {J. Phys. Condens. Matter}\ }\textbf {\bibinfo {volume} {30}},\
  \bibinfo {pages} {485403} (\bibinfo {year} {2018})}\BibitemShut {NoStop}%
\bibitem [{\citenamefont {Ni}\ \emph {et~al.}(2020)\citenamefont {Ni},
  \citenamefont {Xu}, \citenamefont {S{\'a}nchez-Mart{\'i}nez}, \citenamefont
  {Zhang}, \citenamefont {Manna}, \citenamefont {Bernhard}, \citenamefont
  {Venderbos}, \citenamefont {de~Juan}, \citenamefont {Felser}, \citenamefont
  {Grushin},\ and\ \citenamefont {Wu}}]{Ni2020}%
  \BibitemOpen
  \bibfield  {author} {\bibinfo {author} {\bibfnamefont {Z.}~\bibnamefont
  {Ni}}, \bibinfo {author} {\bibfnamefont {B.}~\bibnamefont {Xu}}, \bibinfo
  {author} {\bibfnamefont {M.-{\'A}.}\ \bibnamefont
  {S{\'a}nchez-Mart{\'i}nez}}, \bibinfo {author} {\bibfnamefont
  {Y.}~\bibnamefont {Zhang}}, \bibinfo {author} {\bibfnamefont
  {K.}~\bibnamefont {Manna}}, \bibinfo {author} {\bibfnamefont
  {C.}~\bibnamefont {Bernhard}}, \bibinfo {author} {\bibfnamefont {J.~W.~F.}\
  \bibnamefont {Venderbos}}, \bibinfo {author} {\bibfnamefont {F.}~\bibnamefont
  {de~Juan}}, \bibinfo {author} {\bibfnamefont {C.}~\bibnamefont {Felser}},
  \bibinfo {author} {\bibfnamefont {A.~G.}\ \bibnamefont {Grushin}},\ and\
  \bibinfo {author} {\bibfnamefont {L.}~\bibnamefont {Wu}},\ }\bibfield
  {title} {\bibinfo {title} {{Linear and nonlinear optical responses in the
  chiral multifold semimetal RhSi}},\ }\href
  {https://doi.org/10.1038/s41535-020-00298-y} {\bibfield  {journal} {\bibinfo
  {journal} {npj Quantum Mater.}\ }\textbf {\bibinfo {volume} {5}},\ \bibinfo
  {pages} {96} (\bibinfo {year} {2020})}\BibitemShut {NoStop}%
\bibitem [{\citenamefont {Maslov}\ and\ \citenamefont
  {Chubukov}(2016)}]{Maslov2017}%
  \BibitemOpen
  \bibfield  {author} {\bibinfo {author} {\bibfnamefont {D.~L.}\ \bibnamefont
  {Maslov}}\ and\ \bibinfo {author} {\bibfnamefont {A.~V.}\ \bibnamefont
  {Chubukov}},\ }\bibfield  {title} {\bibinfo {title} {Optical response of
  correlated electron systems},\ }\href
  {https://doi.org/10.1088/1361-6633/80/2/026503} {\bibfield  {journal}
  {\bibinfo  {journal} {Rep. Prog. Phys.}\ }\textbf {\bibinfo {volume} {80}},\
  \bibinfo {pages} {026503} (\bibinfo {year} {2016})}\BibitemShut {NoStop}%
\bibitem [{\citenamefont {Shao}\ \emph {et~al.}(2020)\citenamefont {Shao},
  \citenamefont {Rudenko}, \citenamefont {Hu}, \citenamefont {Sun},
  \citenamefont {Zhu}, \citenamefont {Moon}, \citenamefont {Millis},
  \citenamefont {Yuan}, \citenamefont {Lichtenstein}, \citenamefont {Smirnov},
  \citenamefont {Mao}, \citenamefont {Katsnelson},\ and\ \citenamefont
  {Basov}}]{Shao2020}%
  \BibitemOpen
  \bibfield  {author} {\bibinfo {author} {\bibfnamefont {Y.}~\bibnamefont
  {Shao}}, \bibinfo {author} {\bibfnamefont {A.~N.}\ \bibnamefont {Rudenko}},
  \bibinfo {author} {\bibfnamefont {J.}~\bibnamefont {Hu}}, \bibinfo {author}
  {\bibfnamefont {Z.}~\bibnamefont {Sun}}, \bibinfo {author} {\bibfnamefont
  {Y.}~\bibnamefont {Zhu}}, \bibinfo {author} {\bibfnamefont {S.}~\bibnamefont
  {Moon}}, \bibinfo {author} {\bibfnamefont {A.~J.}\ \bibnamefont {Millis}},
  \bibinfo {author} {\bibfnamefont {S.}~\bibnamefont {Yuan}}, \bibinfo {author}
  {\bibfnamefont {A.~I.}\ \bibnamefont {Lichtenstein}}, \bibinfo {author}
  {\bibfnamefont {D.}~\bibnamefont {Smirnov}}, \bibinfo {author} {\bibfnamefont
  {Z.~Q.}\ \bibnamefont {Mao}}, \bibinfo {author} {\bibfnamefont {M.~I.}\
  \bibnamefont {Katsnelson}},\ and\ \bibinfo {author} {\bibfnamefont {D.~N.}\
  \bibnamefont {Basov}},\ }\bibfield  {title} {\bibinfo {title} {{Electronic
  correlations in nodal-line semimetals}},\ }\href
  {https://doi.org/10.1038/s41567-020-0859-z} {\bibfield  {journal} {\bibinfo
  {journal} {Nat. Phys.}\ }\textbf {\bibinfo {volume} {16}},\ \bibinfo {pages}
  {636} (\bibinfo {year} {2020})}\BibitemShut {NoStop}%
\bibitem [{\citenamefont {Qazilbash}\ \emph {et~al.}(2009)\citenamefont
  {Qazilbash}, \citenamefont {Hamlin}, \citenamefont {Baumbach}, \citenamefont
  {Zhang}, \citenamefont {Singh}, \citenamefont {Maple},\ and\ \citenamefont
  {Basov}}]{Qazilbash2009}%
  \BibitemOpen
  \bibfield  {author} {\bibinfo {author} {\bibfnamefont {M.~M.}\ \bibnamefont
  {Qazilbash}}, \bibinfo {author} {\bibfnamefont {J.~J.}\ \bibnamefont
  {Hamlin}}, \bibinfo {author} {\bibfnamefont {R.~E.}\ \bibnamefont
  {Baumbach}}, \bibinfo {author} {\bibfnamefont {L.}~\bibnamefont {Zhang}},
  \bibinfo {author} {\bibfnamefont {D.~J.}\ \bibnamefont {Singh}}, \bibinfo
  {author} {\bibfnamefont {M.~B.}\ \bibnamefont {Maple}},\ and\ \bibinfo
  {author} {\bibfnamefont {D.~N.}\ \bibnamefont {Basov}},\ }\bibfield  {title}
  {\bibinfo {title} {{Electronic correlations in the iron pnictides}},\ }\href
  {https://doi.org/10.1038/nphys1343} {\bibfield  {journal} {\bibinfo
  {journal} {Nat. Phys.}\ }\textbf {\bibinfo {volume} {5}},\ \bibinfo {pages}
  {647} (\bibinfo {year} {2009})}\BibitemShut {NoStop}%
\bibitem [{\citenamefont {Wenzel}\ \emph {et~al.}(2025)\citenamefont {Wenzel},
  \citenamefont {Uykur}, \citenamefont {Tsirlin}, \citenamefont {Salinas},
  \citenamefont {Ortiz}, \citenamefont {Wilson},\ and\ \citenamefont
  {Dressel}}]{Wenzel2025}%
  \BibitemOpen
  \bibfield  {author} {\bibinfo {author} {\bibfnamefont {M.}~\bibnamefont
  {Wenzel}}, \bibinfo {author} {\bibfnamefont {E.}~\bibnamefont {Uykur}},
  \bibinfo {author} {\bibfnamefont {A.~A.}\ \bibnamefont {Tsirlin}}, \bibinfo
  {author} {\bibfnamefont {A.~N.~C.}\ \bibnamefont {Salinas}}, \bibinfo
  {author} {\bibfnamefont {B.~R.}\ \bibnamefont {Ortiz}}, \bibinfo {author}
  {\bibfnamefont {S.~D.}\ \bibnamefont {Wilson}},\ and\ \bibinfo {author}
  {\bibfnamefont {M.}~\bibnamefont {Dressel}},\ }\bibfield  {title} {\bibinfo
  {title} {{Interplay of $d$- and $p$-states in
  ${\mathrm{RbTi}}_{3}{\mathrm{Bi}}_{5}$ and
  ${\mathrm{CsTi}}_{3}{\mathrm{Bi}}_{5}$ flat-band kagome metals}},\ }\href
  {https://doi.org/10.1103/qq6c-3m2z} {\bibfield  {journal} {\bibinfo
  {journal} {Phys. Rev. B}\ }\textbf {\bibinfo {volume} {112}},\ \bibinfo
  {pages} {L041122} (\bibinfo {year} {2025})}\BibitemShut {NoStop}%
\bibitem [{\citenamefont {Okamura}\ \emph {et~al.}(2020)\citenamefont
  {Okamura}, \citenamefont {Minami}, \citenamefont {Kato}, \citenamefont
  {Fujishiro}, \citenamefont {Kaneko}, \citenamefont {Ikeda}, \citenamefont
  {Muramoto}, \citenamefont {Kaneko}, \citenamefont {Ueda}, \citenamefont
  {Kocsis}, \citenamefont {Kanazawa}, \citenamefont {Taguchi}, \citenamefont
  {Koretsune}, \citenamefont {Fujiwara}, \citenamefont {Tsukazaki},
  \citenamefont {Arita}, \citenamefont {Tokura},\ and\ \citenamefont
  {Takahashi}}]{Okamura2020}%
  \BibitemOpen
  \bibfield  {author} {\bibinfo {author} {\bibfnamefont {Y.}~\bibnamefont
  {Okamura}}, \bibinfo {author} {\bibfnamefont {S.}~\bibnamefont {Minami}},
  \bibinfo {author} {\bibfnamefont {Y.}~\bibnamefont {Kato}}, \bibinfo {author}
  {\bibfnamefont {Y.}~\bibnamefont {Fujishiro}}, \bibinfo {author}
  {\bibfnamefont {Y.}~\bibnamefont {Kaneko}}, \bibinfo {author} {\bibfnamefont
  {J.}~\bibnamefont {Ikeda}}, \bibinfo {author} {\bibfnamefont
  {J.}~\bibnamefont {Muramoto}}, \bibinfo {author} {\bibfnamefont
  {R.}~\bibnamefont {Kaneko}}, \bibinfo {author} {\bibfnamefont
  {K.}~\bibnamefont {Ueda}}, \bibinfo {author} {\bibfnamefont {V.}~\bibnamefont
  {Kocsis}}, \bibinfo {author} {\bibfnamefont {N.}~\bibnamefont {Kanazawa}},
  \bibinfo {author} {\bibfnamefont {Y.}~\bibnamefont {Taguchi}}, \bibinfo
  {author} {\bibfnamefont {T.}~\bibnamefont {Koretsune}}, \bibinfo {author}
  {\bibfnamefont {K.}~\bibnamefont {Fujiwara}}, \bibinfo {author}
  {\bibfnamefont {A.}~\bibnamefont {Tsukazaki}}, \bibinfo {author}
  {\bibfnamefont {R.}~\bibnamefont {Arita}}, \bibinfo {author} {\bibfnamefont
  {Y.}~\bibnamefont {Tokura}},\ and\ \bibinfo {author} {\bibfnamefont
  {Y.}~\bibnamefont {Takahashi}},\ }\bibfield  {title} {\bibinfo {title}
  {{Giant magneto-optical responses in magnetic Weyl semimetal
  Co$_3$Sn$_2$S$_2$}},\ }\href {https://doi.org/10.1038/s41467-020-18470-0}
  {\bibfield  {journal} {\bibinfo  {journal} {Nat. Commun.}\ }\textbf {\bibinfo
  {volume} {11}},\ \bibinfo {pages} {4619} (\bibinfo {year}
  {2020})}\BibitemShut {NoStop}%
\bibitem [{\citenamefont {Stewart}\ \emph {et~al.}(2011)\citenamefont
  {Stewart}, \citenamefont {Yee}, \citenamefont {Liu}, \citenamefont {Kareev},
  \citenamefont {Smith}, \citenamefont {Chapler}, \citenamefont {Varela},
  \citenamefont {Ryan}, \citenamefont {Haule}, \citenamefont {Chakhalian},\
  and\ \citenamefont {Basov}}]{Stewart2011}%
  \BibitemOpen
  \bibfield  {author} {\bibinfo {author} {\bibfnamefont {M.~K.}\ \bibnamefont
  {Stewart}}, \bibinfo {author} {\bibfnamefont {C.-H.}\ \bibnamefont {Yee}},
  \bibinfo {author} {\bibfnamefont {J.}~\bibnamefont {Liu}}, \bibinfo {author}
  {\bibfnamefont {M.}~\bibnamefont {Kareev}}, \bibinfo {author} {\bibfnamefont
  {R.~K.}\ \bibnamefont {Smith}}, \bibinfo {author} {\bibfnamefont {B.~C.}\
  \bibnamefont {Chapler}}, \bibinfo {author} {\bibfnamefont {M.}~\bibnamefont
  {Varela}}, \bibinfo {author} {\bibfnamefont {P.~J.}\ \bibnamefont {Ryan}},
  \bibinfo {author} {\bibfnamefont {K.}~\bibnamefont {Haule}}, \bibinfo
  {author} {\bibfnamefont {J.}~\bibnamefont {Chakhalian}},\ and\ \bibinfo
  {author} {\bibfnamefont {D.~N.}\ \bibnamefont {Basov}},\ }\bibfield  {title}
  {\bibinfo {title} {{Optical study of strained ultrathin films of strongly
  correlated ${\mathrm{LaNiO}}_{3}$}},\ }\href
  {https://doi.org/10.1103/PhysRevB.83.075125} {\bibfield  {journal} {\bibinfo
  {journal} {Phys. Rev. B}\ }\textbf {\bibinfo {volume} {83}},\ \bibinfo
  {pages} {075125} (\bibinfo {year} {2011})}\BibitemShut {NoStop}%
\bibitem [{\citenamefont {Wenzel}\ \emph {et~al.}(2024)\citenamefont {Wenzel},
  \citenamefont {Uykur}, \citenamefont {Tsirlin}, \citenamefont {Pal},
  \citenamefont {Roy}, \citenamefont {Yi}, \citenamefont {Shekhar},
  \citenamefont {Felser}, \citenamefont {Pronin},\ and\ \citenamefont
  {Dressel}}]{Wenzel2024}%
  \BibitemOpen
  \bibfield  {author} {\bibinfo {author} {\bibfnamefont {M.}~\bibnamefont
  {Wenzel}}, \bibinfo {author} {\bibfnamefont {E.}~\bibnamefont {Uykur}},
  \bibinfo {author} {\bibfnamefont {A.~A.}\ \bibnamefont {Tsirlin}}, \bibinfo
  {author} {\bibfnamefont {S.}~\bibnamefont {Pal}}, \bibinfo {author}
  {\bibfnamefont {R.~M.}\ \bibnamefont {Roy}}, \bibinfo {author} {\bibfnamefont
  {C.}~\bibnamefont {Yi}}, \bibinfo {author} {\bibfnamefont {C.}~\bibnamefont
  {Shekhar}}, \bibinfo {author} {\bibfnamefont {C.}~\bibnamefont {Felser}},
  \bibinfo {author} {\bibfnamefont {A.~V.}\ \bibnamefont {Pronin}},\ and\
  \bibinfo {author} {\bibfnamefont {M.}~\bibnamefont {Dressel}},\ }\bibfield
  {title} {\bibinfo {title} {{Intriguing Low-Temperature Phase in the
  Antiferromagnetic Kagome Metal FeGe}},\ }\href
  {https://doi.org/10.1103/PhysRevLett.132.266505} {\bibfield  {journal}
  {\bibinfo  {journal} {Phys. Rev. Lett.}\ }\textbf {\bibinfo {volume} {132}},\
  \bibinfo {pages} {266505} (\bibinfo {year} {2024})}\BibitemShut {NoStop}%
\bibitem [{\citenamefont {Miranda}\ \emph {et~al.}(2022)\citenamefont
  {Miranda}, \citenamefont {Klautau}, \citenamefont {Bergman},\ and\
  \citenamefont {Petrilli}}]{Miranda2022}%
  \BibitemOpen
  \bibfield  {author} {\bibinfo {author} {\bibfnamefont {I.~P.}\ \bibnamefont
  {Miranda}}, \bibinfo {author} {\bibfnamefont {A.~B.}\ \bibnamefont
  {Klautau}}, \bibinfo {author} {\bibfnamefont {A.}~\bibnamefont {Bergman}},\
  and\ \bibinfo {author} {\bibfnamefont {H.~M.}\ \bibnamefont {Petrilli}},\
  }\bibfield  {title} {\bibinfo {title} {{Band filling effects on the emergence
  of magnetic skyrmions: Pd/Fe and Pd/Co bilayers on Ir(111)}},\ }\href
  {https://doi.org/10.1103/PhysRevB.105.224413} {\bibfield  {journal} {\bibinfo
   {journal} {Phys. Rev. B}\ }\textbf {\bibinfo {volume} {105}},\ \bibinfo
  {pages} {224413} (\bibinfo {year} {2022})}\BibitemShut {NoStop}%
\bibitem [{\citenamefont {Dup{\'e}}\ \emph {et~al.}(2016)\citenamefont
  {Dup{\'e}}, \citenamefont {Bihlmayer}, \citenamefont {B{\"o}ttcher},
  \citenamefont {Bl{\"u}gel},\ and\ \citenamefont {Heinze}}]{Dupe2016}%
  \BibitemOpen
  \bibfield  {author} {\bibinfo {author} {\bibfnamefont {B.}~\bibnamefont
  {Dup{\'e}}}, \bibinfo {author} {\bibfnamefont {G.}~\bibnamefont {Bihlmayer}},
  \bibinfo {author} {\bibfnamefont {M.}~\bibnamefont {B{\"o}ttcher}}, \bibinfo
  {author} {\bibfnamefont {S.}~\bibnamefont {Bl{\"u}gel}},\ and\ \bibinfo
  {author} {\bibfnamefont {S.}~\bibnamefont {Heinze}},\ }\bibfield  {title}
  {\bibinfo {title} {{Engineering skyrmions in transition-metal multilayers for
  spintronics}},\ }\href {https://doi.org/10.1038/ncomms11779} {\bibfield
  {journal} {\bibinfo  {journal} {Nat. Commun.}\ }\textbf {\bibinfo {volume}
  {7}},\ \bibinfo {pages} {11779} (\bibinfo {year} {2016})}\BibitemShut
  {NoStop}%
\bibitem [{\citenamefont {Markiewicz}(1991)}]{Markiewicz1991}%
  \BibitemOpen
  \bibfield  {author} {\bibinfo {author} {\bibfnamefont {R.~S.}\ \bibnamefont
  {Markiewicz}},\ }\bibfield  {title} {\bibinfo {title} {{Van Hove singularity
  and high-$T_c$ superconductivity: a review}},\ }\href
  {https://doi.org/10.1142/S0217979291000791} {\bibfield  {journal} {\bibinfo
  {journal} {Int. J. Mod. Phys. B}\ }\textbf {\bibinfo {volume} {05}},\
  \bibinfo {pages} {2037} (\bibinfo {year} {1991})}\BibitemShut {NoStop}%
\bibitem [{\citenamefont {Irkhin}\ \emph {et~al.}(2002)\citenamefont {Irkhin},
  \citenamefont {Katanin},\ and\ \citenamefont {Katsnelson}}]{Irkhin2002}%
  \BibitemOpen
  \bibfield  {author} {\bibinfo {author} {\bibfnamefont {V.~Y.}\ \bibnamefont
  {Irkhin}}, \bibinfo {author} {\bibfnamefont {A.~A.}\ \bibnamefont
  {Katanin}},\ and\ \bibinfo {author} {\bibfnamefont {M.~I.}\ \bibnamefont
  {Katsnelson}},\ }\bibfield  {title} {\bibinfo {title} {{Robustness of the Van
  Hove Scenario for High-${T}_{c}$ Superconductors}},\ }\href
  {https://doi.org/10.1103/PhysRevLett.89.076401} {\bibfield  {journal}
  {\bibinfo  {journal} {Phys. Rev. Lett.}\ }\textbf {\bibinfo {volume} {89}},\
  \bibinfo {pages} {076401} (\bibinfo {year} {2002})}\BibitemShut {NoStop}%
\end{thebibliography}%

\end{document}


\widetext
\begin{center}
\textbf{\large Supplemental Material for ''Electronic structure of the Gd-based intermetallics \GCG\ and \GCA''}

\vspace{5mm}
M. Pinteri\'{c}, M. Dressel, P. Puphal, and M. Wenzel
\end{center}
\setcounter{equation}{0}
\setcounter{figure}{0}
\setcounter{table}{0}
\setcounter{page}{1}
\makeatletter
\renewcommand{\theequation}{S\arabic{equation}}
\renewcommand{\thefigure}{S\arabic{figure}}
\renewcommand{\bibnumfmt}[1]{[S#1]}
\renewcommand{\citenumfont}[1]{S#1}

\section{Decomposition at selected temperatures}
\begin{figure}[!h]
	\centering
	\includegraphics[width=0.68\columnwidth]{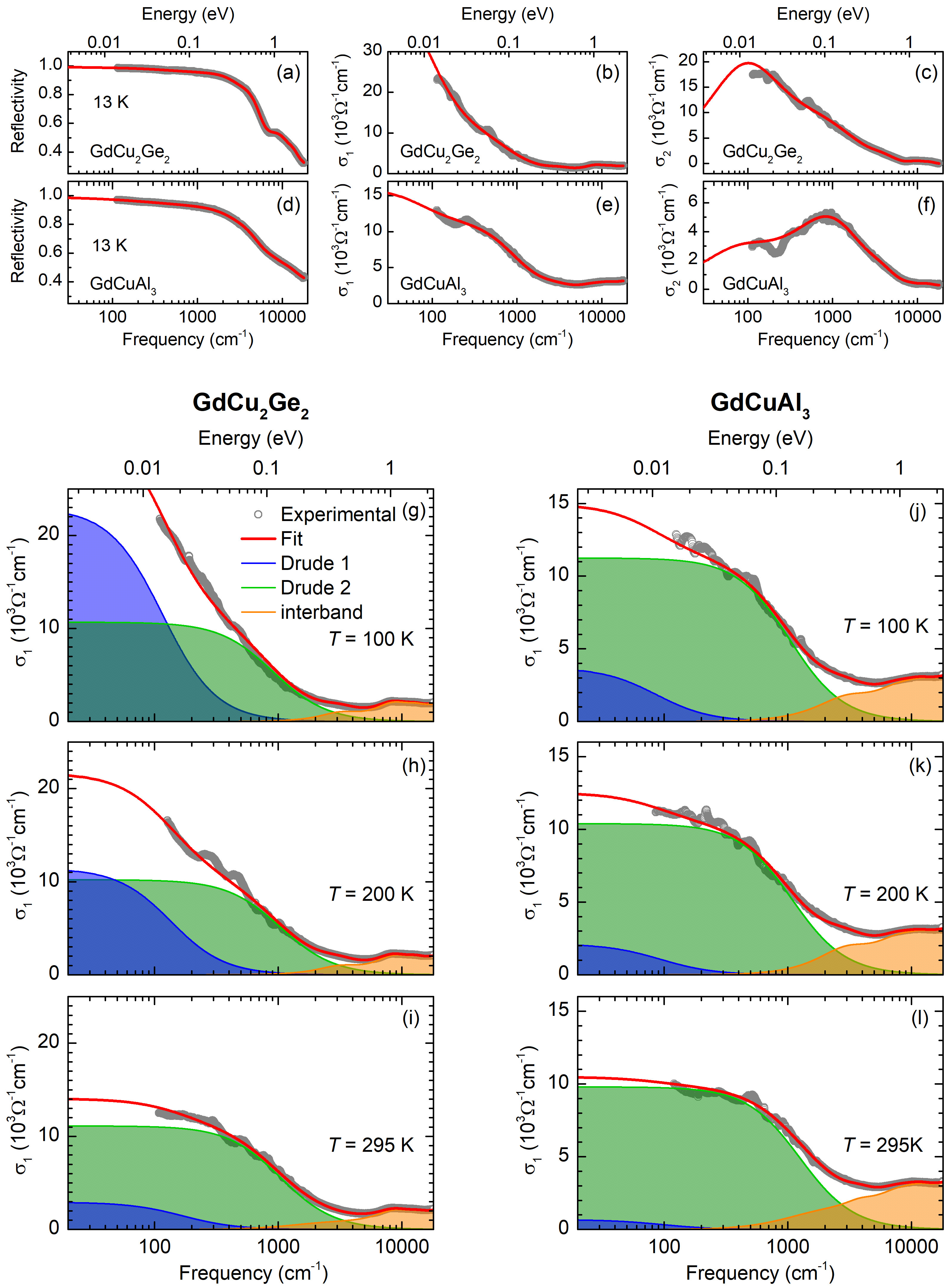}
	\caption{Modeled reflectivity (a), real part of the optical conductivity (b), and imaginary part of the optical conductivity (c) of \GCG\ at 13~K. Modeled reflectivity (d), real part of the optical conductivity (e), and imaginary part of the optical conductivity (f) of \GCA\ at 13~K. Decomposed real part of the optical conductivity of \GCG~(g-i) and \GCA~(j-l) at 100~K, 200~K, and 295~K. The data are modeled using two Drude contributions (blue and green) and five Lorentzians above 2000~\cm, describing the interband transitions~(orange). }
	\label{decomp_SM}
\end{figure}

In Fig.~\ref{decomp_SM} we present the fitted optical spectra at 13~K and decomposed optical conductivities of \GCG\ and \GCA\ at 100~K, 200~K, and 295~K modeled by the Drude-Lorentz approach discussed in the main text. 
\clearpage
\section{Additional computational results}
\begin{figure}[h]
	\centering
	\includegraphics[width=0.7\columnwidth]{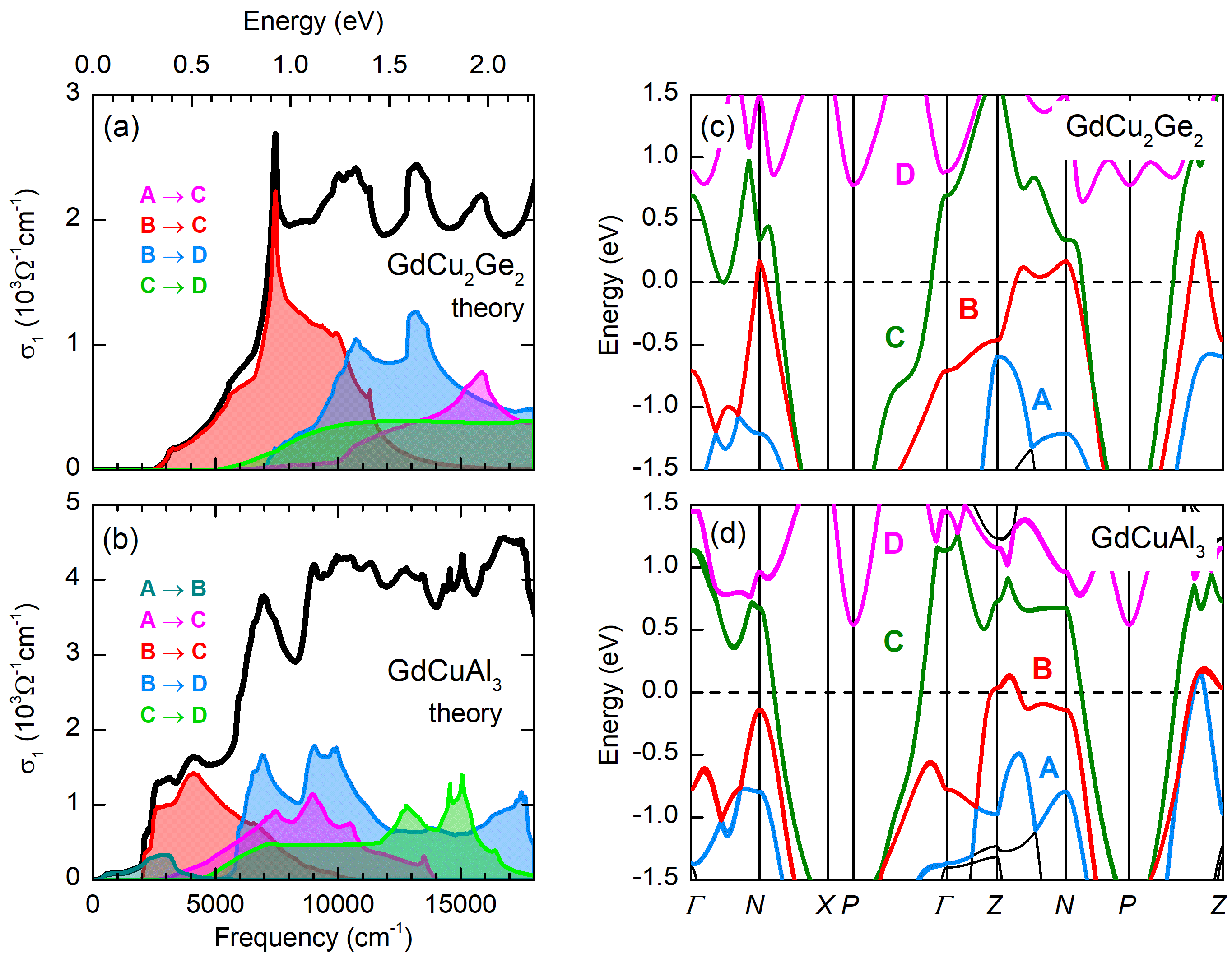}
	\caption{(a, b) Calculated band-resolved optical conductivities with different colors marking interband transitions across different bands labeled in the band structures in panels (c) and (d). Note that no broadening was applied to these spectra and the calculated energies of the optical conductivity of \GCA\ are rescaled by a factor of 1.5 to match the experimentally observed energies. }		
	\label{bandresolved}
\end{figure}

Fig.~\ref{bandresolved} displays the band-resolved optical conductivities. Consistent with the highly similar band structures, the interband optical transitions are also very similar in both compounds, with comparable contributions from the relevant bands. In \GCG, the interband absorption below $\sim$0.7~eV arises exclusively from transitions between bands 'B' and 'C', following the labeling in panel~(c). In \GCA, transitions between bands 'A' and 'B' provide an additional, albeit minor, contribution because band 'A' is shifted upward by hybridization with the Al 3$p$ states and crosses the Fermi level along the $P$--$Z$ direction. 

\begin{figure}[h]
	\centering
	\includegraphics[width=0.9\columnwidth]{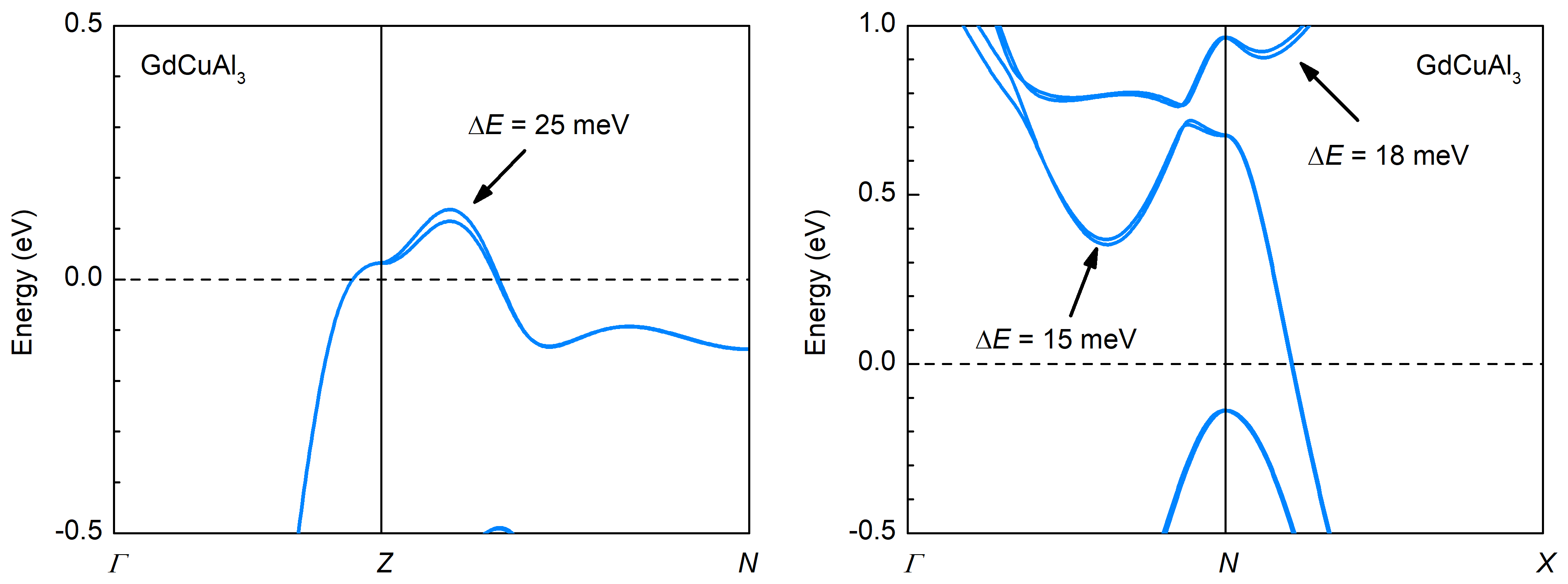}
	\caption{ Spin-orbit-coupling-induced band splittings along selected $k$-directions in \GCA.}
	\label{splitting}
\end{figure}

Fig.~\ref{splitting} offers a zoomed-in view of the band structure of \GCA, highlighting the spin-orbit-coupling-induced band splittings on the order of 20~meV along selected $k$-directions.

As Gd has a half-filled 4$f$ shell, these states are highly localized and do not contribute to the low-energy electronic structure of Gd-based compounds. Density functional theory offers several approaches to account for the highly localized nature of $f$ electrons. In the main text, we employ the open-core approximation, which allows us to model the system without magnetic ordering, consistent with the paramagnetic phase in the temperature range relevant to our study.

\begin{figure}
	\centering
	\includegraphics[width=0.9\columnwidth]{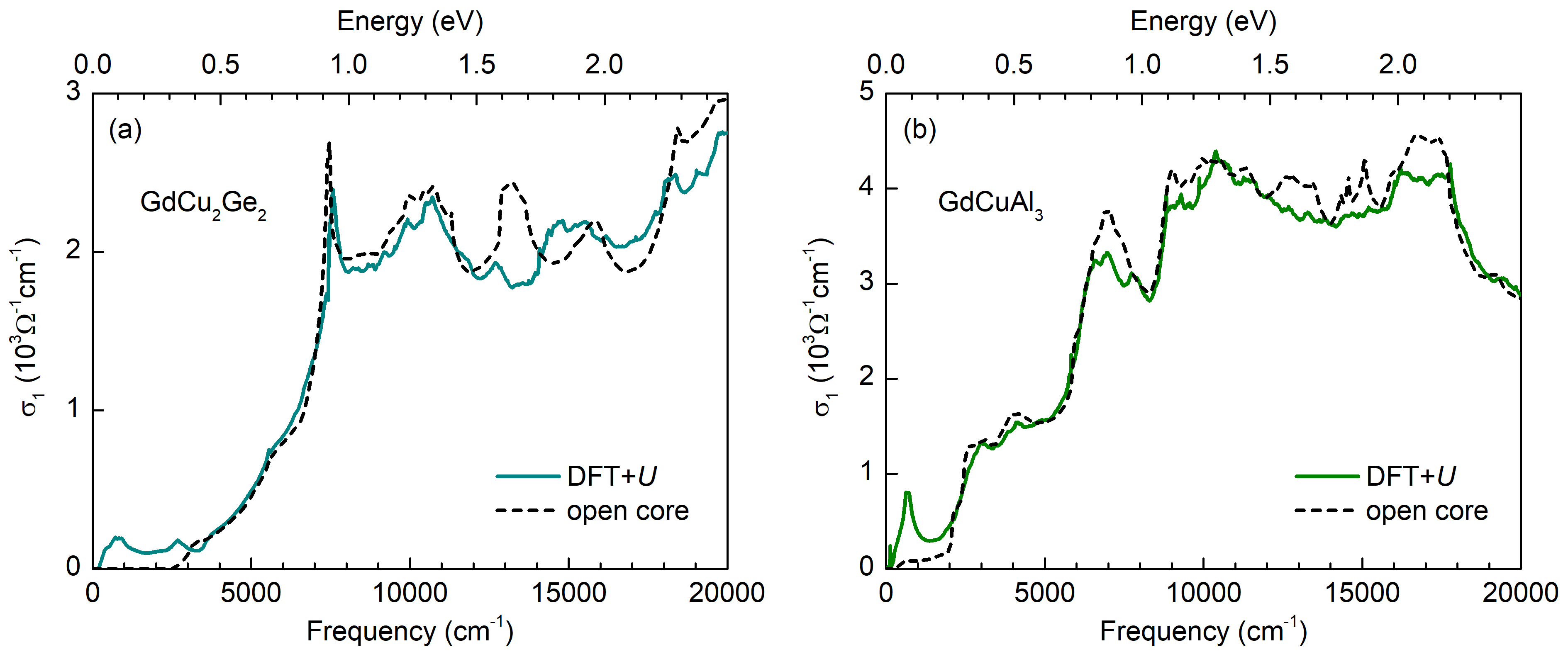}
	\caption{Theoretical optical conductivities of \GCG\ (a) and \GCA\ (b) calculated with the open-core approximation (dashed lines) and the DFT+$U$ approach (colored solid lines). }
	\label{OCcomp}
\end{figure}

An alternative approach is the DFT+$U$ method, in which a large Hubbard $U$ is applied to the 4$f$ electrons, effectively localizing them and shifting the occupied and unoccupied states well below and above the Fermi level, respectively. In this case, however, spin polarization must be included to obtain convergence to the Gd$^{3+}$ configuration. For our calculations, we impose an in-plane antiferromagnetic ordering in which each Gd layer is ferromagnetically ordered, while adjacent layers are coupled antiferromagnetically. Because the Gd 5$d$ states contribute significantly to the low-energy electronic structure, the resulting band structure is affected by this magnetic ordering. Nevertheless, the calculated optical conductivity remains comparable to that obtained within the open-core approximation, as shown in Fig.~\ref{OCcomp}.

Moreover, the DFT-derived plasma frequencies, and consequently, the ratio $\omega_{\mathrm{p, exp}}^2/\omega_{\mathrm{p, DFT}}^2$ used as a gauge of electronic correlations, remain largely unaffected by the different treatment of the Gd 4$f$ states as outline in Tab.~\ref{tableI}. 

\begin{table}
\centering
\caption{Comparison of the ratio $\omega_{\mathrm{p, exp}}^2/\omega_{\mathrm{p, DFT}}^2$ used as a gauge of electronic correlations, derived from the open-core approximation and DFT+$U$ at 13~K.}
\label{tableI}
\begin{tabular*}{\columnwidth}{@{\extracolsep{\fill}}lccccc@{}}
\toprule
& $\omega_{\mathrm{p, exp}}$ (\cm) & $\omega_{\mathrm{p, open-core}}$ (\cm) &  $\omega_{\mathrm{p, exp}}^2/\omega_{\mathrm{p, open-core}}^2$ & $\omega_{\mathrm{p, DFT+}U}$ (\cm) &$\omega_{\mathrm{p, exp}}^2/\omega_{\mathrm{p, DFT+}U}^2$ \\
\midrule
\GCG & 27098 & 39733 &  0.47 & 39221 &0.48 \\
\GCA & 26684 & 39059 &  0.47 &37782 & 0.50 \\
\bottomrule
\end{tabular*}
\end{table}